# Real-space Visualization of Emergent Electron Crystals in Rhombohedral Graphene

Yiming Sun[1,2*], Jinghao Deng[1*†], Jiabin Xie[1], Donghan Ge[1], Hongyuan Li[1,3], Takashi Taniguchi[4], Kenji Watanabe[5], Xiaomeng Liu[1†]

1 *Laboratory of Atomic and Solid State Physics, Cornell University, Ithaca, NY, USA*

2 *School of Applied and Engineering Physics, Cornell University, Ithaca, NY, USA*

3 *Kavli Institute at Cornell for Nanoscale Science, Ithaca, NY, USA*

4 *Research Center for Materials Nanoarchitectonics, National Institute for Materials Science, Tsukuba, Japan*

5 *Research Center for Electronic and Optical Materials, National Institute for Materials Science, Tsukuba, Japan*

* *These authors contributed equally to this work*

† *Correspondence and requests for materials should be addressed to Xiaomeng Liu (xl926@cornell.edu) or Jinghao Deng (jd2353@cornell.edu)*

**Strongly interacting electrons can spontaneously break spatial symmetries to form electron crystals, exemplified by the Wigner crystal. Recent theoretical and experimental studies of topological flat bands in rhombohedral graphene have suggested the possibility of more exotic forms of electronic crystallization, including anomalous Hall crystals that entangle charge order with nontrivial topology[1–6], and metallic electron crystals in which localized and itinerant carriers coexist[7–9]. Direct real-space observation of these unconventional crystalline states, however, has remained elusive. Here we use scanning tunneling microscopy and spectroscopy to visualize emergent electronic crystals in rhombohedral hexalayer graphene. At low electric fields and over a finite range of hole doping, we observe electronic lattice patterns that evolve from a honeycomb structure to an oblique structure through a first-order quantum phase transition with increasing hole density. The Fermi surface extracted from quasiparticle-interference measurements lacks the geometry to account for the observed crystal patterns through conventional nesting. Together with metallic transport and a crystal-site density much lower than the doped carrier density, this supports metallic electron crystals in which a subset of the carriers crystallizes. The honeycomb crystal occupies the same phase space with the multiferroic orbital magnetism observed in previous transport experiments[10] and exhibits domain stabilization by a small magnetic field, which may suggest a metallic anomalous Hall crystal. With increasing magnetic fields, the oblique phase develops a $\sqrt{2} \times \sqrt{2}$ reconstruction with a crystal sublattice energy splitting that increases linearly with field, corresponding to a *g*-factor of 16. It could be interpreted as an orbital-antiferromagnetic electron crystal with alternating orbital magnetization across the crystal lattices. These results establish a new paradigm of electron crystallization in which charge order is intertwined with orbital magnetism.**

Unlike atoms, which commonly crystallize upon cooling, large collections of electrons typically remain in liquid states down to zero temperature because of their quantum nature. Only in rare regimes where long-range Coulomb interactions overwhelm kinetic energy do electrons localize and form a quantum solid. The canonical example is the Wigner crystal[11], in which electrons arrange into a triangular lattice in two dimensions. It has been realized in electrons on helium[12], partially filled Landau levels[13–15], parabolic bands at low densities[16–22], and moiré semiconductors[23–26], where kinetic energy is strongly reduced or quenched. In these settings, electron crystallization is typically associated with insulating behavior and with charge patterns that are either triangular or constrained by an underlying moiré lattice.

Topological flat bands in rhombohedral multilayer graphene provide a fundamentally different setting for electron crystallization. Their electrically tunable flat bands combine strong Coulomb interactions with large valley-contrasting Berry curvature[27–29], giving rise to a rapidly expanding family of correlated and topological phases, including signatures of chiral superconductivity[30], orbital magnetism[31], and integer and fractional Chern insulators[32–35]. Growing theoretical and experimental evidence suggests that real-space electronic order may underlie many of these phenomena[1–8,36–39]. For example, anomalous Hall crystals, which spontaneously break translational symmetry while carrying a quantized Hall response, have been proposed as a mechanism for interaction-driven Chern bands[1,2,4,5,40], while hysteretic current–voltage characteristics observed in transport suggest possible crystallization in certain insulating and metallic states[8,33,38]. These developments suggest that electron crystallization in topological flat bands may extend beyond the conventional Wigner paradigm, potentially giving rise to electron crystals with unconventional lattice symmetry, metallicity and nontrivial topology. Determining the real-space structure of these proposed electron crystals is therefore essential for understanding the microscopic origin of the correlated and topological phases in rhombohedral graphene. Yet, definitive evidence of spatial orders in this system has remained absent.

Real-space imaging provides the most direct evidence of electron crystallization while revealing its lattice symmetry, periodicity, domain structure, and microscopic organization. 1D Wigner crystal has been visualized in carbon nanotubes[41] and along transition metal dichalcogenide (TMD) domain walls[42] using scanning-probe techniques. In two dimensions, scanning tunnelling microscopy has imaged Wigner crystals in TMD mono-/bi-layers[20–22], transition metal trihalides[43,44], and quantum Hall systems[45,46], as well as generalized Wigner crystals in moiré heterostructures[23,26,47]. These electron crystals typically exhibit triangular or distorted triangular lattices, or adopt geometries constrained by an underlying atomic or moiré lattice. None has revealed nontrivial internal structure beyond its charge order. Although Wigner crystals in quantum Hall systems form within a topological electronic background established by the filled Landau levels, the crystalline order itself is conventionally understood as a topologically trivial Wigner solid. Direct real-space observation of a spontaneously formed electron crystal exhibiting intrinsic orbital magnetism—and potentially nontrivial topology—has therefore remained elusive.

Here, using scanning tunnelling microscopy and spectroscopy (STM/STS), we directly visualize spontaneous electron crystallization in rhombohedral hexalayer graphene (R6G). We uncover a transition from honeycomb to oblique crystal lattices with increasing hole doping, accompanied by hysteresis and phase coexistence at the phase boundary. Both crystals emerge in a metallic regime, and their crystal-site densities are approximately an order of magnitude lower than the doped-hole density, distinguishing them from conventional Wigner crystals. The honeycomb crystal coincides in the phase diagram with an orbital-magnetic state characterized by an anomalous Hall response and exhibits domains that can be aligned by a small perpendicular magnetic field. These observations suggest that the honeycomb phase may be a metallic anomalous Hall crystal. By contrast, the magnetic-field response of the oblique crystal is consistent with an orbital-antiferromagnetic state, in which the spontaneously formed electronic lattice hosts oppositely oriented orbital magnetic moments on neighboring sites. These observations reveal a new regime of electron crystallization in which charge order exhibits unconventional symmetry and is intimately connected to orbital magnetism.

## Spectroscopic properties of rhombohedral hexalayer graphene

Our STM experiments were performed on an R6G device with a graphite back gate (Fig. 1a). The back-gate voltage $V_{\mathrm{G}}$ simultaneously controls the carrier density, $n = n_0 + \frac{\varepsilon V_{\mathrm{G}}}{d}$, and the perpendicular electric field, $E = E_0 + \frac{V_{\mathrm{G}}}{2d}$, where $n_0$ and $E_0$ are the built-in doping and electric field, $d$ is the hBN dielectric thickness and $\varepsilon$ is the dielectric constant. Because the device does not include a top gate, the experiment probes along a linear trajectory through the two-dimensional $n$-$E$ phase space.

The gate-dependent tunneling spectra in Fig. 1f exhibit pronounced fluctuations in the differential tunneling conductance ($\mathrm{d}I/\mathrm{d}V$) over the hole-doped gate-voltage range between the orange and green dashed lines. These fluctuations are consistent across repeated measurements and are therefore intrinsic rather than measurement artifacts (Extended Data Fig. 1). As demonstrated below through spatially resolved measurements, they are associated with the formation of electron crystal phases. Previous transport measurements on a dual-gated device revealed strongly fluctuating resistance (Extended Data Fig. 2) within a triangular region of the $n$-$E$ phase space[10]. This region likely corresponds to the electron crystal phases responsible for the spectral fluctuations observed here. Based on this correspondence and the evolution of other band features, we estimate that the gate-voltage trajectory accessed in our experiment follows the black line in Fig. 1e.

The point spectra also reveal an insulating gap near $V_{\mathrm{G}} \approx 0$ V, consistent with the correlated insulating state shown in Fig. 1e. Previous studies[10] of the electronic structure have established that doping this correlated insulator produces flavor-split bands with opposite layer polarization. For $V_{\mathrm{G}} < -1$ V, the spectra exhibit three prominent peaks: one below, one near, and one above the Fermi energy. We assign the lower- and middle-energy peaks to the van Hove singularities of the top- and bottom-layer valence bands, respectively, while the highest-energy peak likely originates from the conduction band (see Extended Data Fig. 3 for detailed discussion). Spectral features associated with the top layer are considerably stronger because they couple directly to the STM tip, whereas states from the bottom layer are attenuated by their greater distance from the surface. A detailed analysis of electronic structures, including spectral features observed in Fig. 1f, is provided in Extended Data Fig. 3.

To complement the spectral measurements, we performed in situ Corbino transport by poking the STM tip into the sample (see Methods for details). The resulting conductance trace (blue curve in Fig. 1f) confirms the correlated insulating state near $V_{\mathrm{G}} \approx 0$ V and metallic behavior away from charge neutrality. A change in conductance is observed at the yellow dashed line, marking the transition between two crystal phases.

To establish the band structure from which the electron crystal emerges, we performed quasiparticle interference (QPI) measurements near isolated defects. Figure 1b shows the QPI dispersion at $V_{\mathrm{G}} = -1$ V, just outside the electron-crystal phase boundary, obtained by Fourier transforming spectroscopic measurements acquired along a linecut across a defect. A pronounced Mexican-hat-shaped valence band is observed, while weaker features above approximately 5 mV are attributed to the conduction band. At this gate voltage, the perpendicular electric field is small, and the bands remain approximately layer degenerate. Two-dimensional QPI measured at a bias of 1 mV (Fig. 1c) reveals a pronounced trigonal warping effect near the valence-band edge, whereas the QPI at -2 mV (Fig. 1d) displays the expected annular contour. Across the full range of gate and bias voltages investigated, all QPI measurements exhibit either circular or threefold symmetric ($C_3$) scattering patterns, with no indication of spontaneous rotational symmetry breaking in the momentum-space electronic structure (Fig. 1c, d; 2a, c, g).

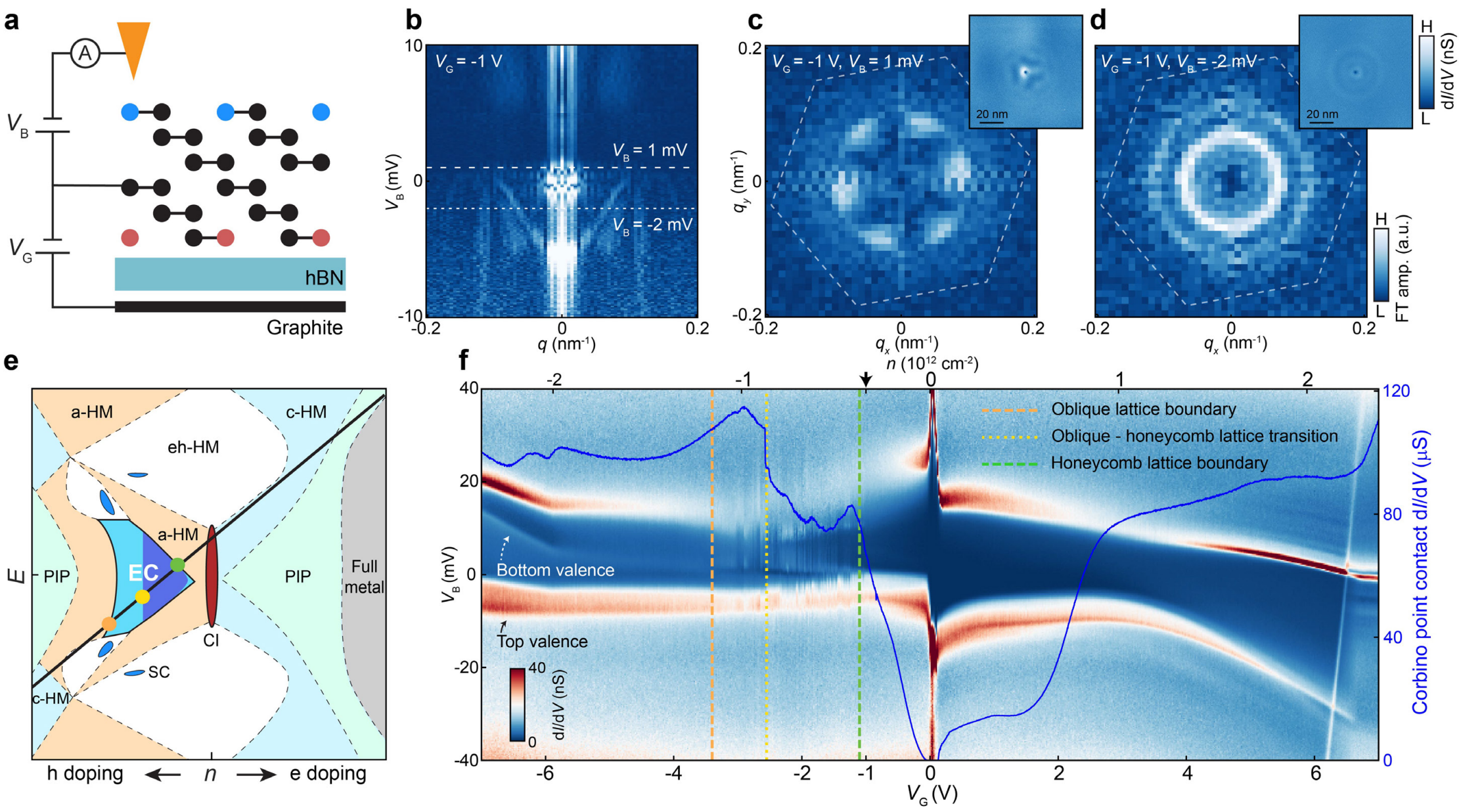

**Fig. 1 | Gate-dependent spectra, QPI dispersion, and QPI images of rhombohedral hexalayer graphene (R6G). a**, Schematic of STM experiment on R6G. Blue and red spheres denote the non-dimerized sites on the top and bottom layers, which dominate the low-energy electronic structure of R6G. **b**, QPI dispersion at $V_{\mathrm{G}}$ = -1 V, obtained by Fourier transforming a 1D spectroscopic linecut acquired across an isolated defect (details in Methods). The characteristic Mexican-hat-shaped dispersion is clearly resolved. Setpoint parameters for hole excitations: $V_{\mathrm{B}}$ = 100 mV, $I_{\mathrm{t}}$ = 100 pA, $\Delta z$ = -

165 pm (see Methods); for electron excitations: $V_B$ = 1 V, $I_t$ = 100 pA, Δ$z$ = -390 pm. **c, d,** Two-dimensional $q$-space QPI maps at representative bias voltages of $V_B$ = 1 mV and -2 mV, with real-space QPI images shown in insets. Gaussian masks are applied to suppress the low frequency background (see Methods). R6G Bragg vector directions are indicated by the vertices of the white dashed hexagons. H stands for high and L stands for low. Setpoint parameters: $V_B$ = 100 mV, $I_t$ = 100 pA, Δ$z$ = -160 pm. **e,** Schematic phase diagram of R6G derived from transport measurements in previous work[10], with the inferred trajectory of the single-gate STM measurements indicated by the black line. EC, electron crystal; CI, correlated insulator; a-HM, annular half metal; c-HM, circular half metal; eh-HM, electron-hole half metal; SC, superconductivity; PIP, partial isospin polarized state. **f**, Gate-dependent point tunneling spectra d$I$/d$V$($V_G$, $V_B$) (setpoint parameters: $V_B$ = 100 mV, $I_t$ = 1 nA, Δ$z$ = -20 pm) and two-terminal Corbino conductance measured *in situ* after STS measurements (blue curve, right axis; see Methods). The phase boundaries of the oblique and honeycomb crystals, determined from spectroscopic imaging (Extended Data Fig. 6), are indicated by dashed lines and by corresponding circular markers in **e**. Layer indices (top/bottom) of prominent spectroscopic features are marked on the left. Gate voltage in **b-d** is highlighted with a black arrow on the top axis.

**Observation of oblique and honeycomb electron crystals**

Within the gate-voltage range bounded by the orange and green lines in Fig. 1f, spectroscopic imaging near the Fermi energy reveals a sequence of stripe, oblique, and honeycomb patterns. Outside this regime, at gate voltage $V_G$ = −3.8 V, the spectroscopic image shows only QPI at the Fermi energy (Fig. 2b), characterized by an approximately circular scattering contour in Fourier space (Fig. 2a). Upon crossing the orange phase boundary to $V_G$ = −3.4 V, an additional stripe-like modulation emerges alongside the QPI pattern (Fig. 2c). Its Fourier transform exhibits two peaks oriented approximately along $q_x$, with a wavevector length substantially smaller than those of the dominant QPI features.

Moving past this transient stripe regime, the long-wavelength modulation develops into a near-square oblique lattice, as shown at $V_G$ = −2.68 V in Fig. 2d. Its Fourier transform exhibits peaks corresponding to two primitive reciprocal-lattice vectors: one is nearly aligned with a crystallographic direction of R6G, whereas the other is not aligned with any high-symmetry direction (Fig. 2d inset). The resulting pattern explicitly breaks the threefold rotational symmetry of rhombohedral graphene and shows an unexpectedly large real-space period of approximately 30 nm. The modulation amplitude is also pronounced, with the d$I$/d$V$ signal differing by more than a factor of two between the maxima and minima (Extended Data Fig. 4).

Further increasing the gate voltage causes an abrupt reorientation of the oblique lattice by approximately 20°, while leaving its period nearly unchanged (Fig. 2e). The rotated state persists over a narrow gate voltage interval of 0.04 V before suddenly transitioning into a honeycomb pattern (Fig. 2f). The density of states is enhanced at the two sublattices of the honeycomb pattern, with a slight contrast between them. For most of the honeycomb regime, the reciprocal-lattice vectors align with the crystallographic directions of R6G (Fig. 2f, inset), but deviate substantially from them near the green phase boundary as discussed below. Larger-area images reveal spatial variations in the visibility of the honeycomb modulation, hinting at the presence of domains (Extended Data Fig. 5). Upon further increasing the gate voltage to $V_G = -1$ V, the crystalline modulation disappears and the system returns to a state characterized by QPI (Fig. 2g, h). Throughout the crystalline regime, the crystal wavevectors' magnitudes remain substantially smaller than the dominant QPI wavevectors.

We next quantify the crystal's density, symmetry, and orientation through its evolution in gate voltage by Fourier transforming each spectroscopic image (see Extended Data Fig. 6, 8) and converting the resulting crystal wavevectors into real-space lattice vectors. From these, we determine the crystal unit cell density $n_{Crystal}$, the inter-vector angle $\theta_{Crystal}$, and the misalignment angle $\theta_{Crystal-R6G}$ between the crystal and R6G lattices (Methods).

Throughout the crystalline regime, the crystal density is approximately an order of magnitude lower than the gate-induced hole density (Fig. 2i), inconsistent with a conventional Wigner crystal containing one doped carrier per primitive unit cell. Instead, only a minority of the doped carriers crystallize, while the remainder stays itinerant. This is consistent with the coexistence of QPI and a crystalline pattern with different length scales across the crystal regime, particularly notable for spectroscopic images at the Fermi energy (Fig. 2c and Extended Data Fig. 7). Lastly, measurements at 0.3 K and 1.6 K yield consistent crystal densities, which decrease towards zero as the green melting boundary is approached.

The symmetry change is captured by inter-vector angle $\theta_{Crystal}$ (Fig. 2j), which would be 90° for a square Bravais lattice and 60° for a triangular Bravais lattice. It remains near 80-85° throughout the oblique phase before jumping to approximately 60° upon entering the honeycomb phase. Notably, $n_{Crystal}$ evolves continuously across the oblique-to-honeycomb transition despite the abrupt change in lattice symmetry. This behavior suggests that a well-defined population of carriers crystallizes and reorganizes between oblique and honeycomb configurations, favoring an electron crystal interpretation over a conventional charge-density-wave interpretation.

The crystal orientation, quantified by $\theta_{Crystal-R6G}$, is shown in Fig. 2k. The abrupt 20° rotation of the oblique lattice near the oblique-to-honeycomb phase boundary is evident. This rotation is hysteretic: forward and backward gate sweeps yield different crystal orientations at the same gate voltage, as highlighted by the purple bar in Fig. 2k inset (corresponding images shown in Extended Data Fig. 6b, c). Although $\theta_{Crystal-R6G}$ stays near zero throughout most of the crystalline regime, the crystal orientation deviates markedly from the R6G crystallographic axes near both the orange and green phase boundaries.

The abrupt reorientation, together with its gate-voltage hysteresis, indicates a first-order transition between two orientationally distinct oblique states. Separately, at the oblique-to-honeycomb transition, the two lattice symmetries coexist as spatially separated domains (Fig. 2l and Extended Data Fig. 5). A well-defined boundary separates the oblique and honeycomb regions, with their distinct reciprocal-space structures resolved by Fourier transforms of the lower and upper domains, respectively (Fig. 2l, insets). The abrupt change in lattice symmetry and real-space phase coexistence provides evidence that the oblique-to-honeycomb transition is also first order. Moreover, the gate-dependent crystal orientation, unconventional crystal symmetries, large modulation amplitude, and mismatch between the crystal and QPI wavevectors are inconsistent with a conventional weak-coupling charge-density wave driven by Fermi-surface nesting.

Finally, energy-resolved spectroscopy reveals the spatially varying electronic structure of the two crystal phases. Figure 2m, n show spectroscopic linecuts through high-symmetry points of the oblique and honeycomb lattices, respectively. In the oblique phase, the spatial modulation is strongest for the middle spectroscopic peak near $V_B$ = 2 mV (Fig. 2m), consistent with the bottom-layer-polarized valence band dominating the observed crystal contrast. The higher-energy peak near $V_B$ = 8 mV modulates in phase with the middle peak, whereas the lower-energy peak near $V_B$ = -5 mV modulates out of phase.

A similar relationship among the three spectroscopic features is observed in the honeycomb phase (Fig. 2n). Notably, the two sites of the honeycomb basis exhibit distinct local spectra: site A exhibits a broader peak at higher energy, whereas site B displays a sharper peak at lower energy. The inequivalent local electronic structures of the two basis sites in the honeycomb lattice are consistent with the theoretical description of anomalous Hall crystals[1,5].

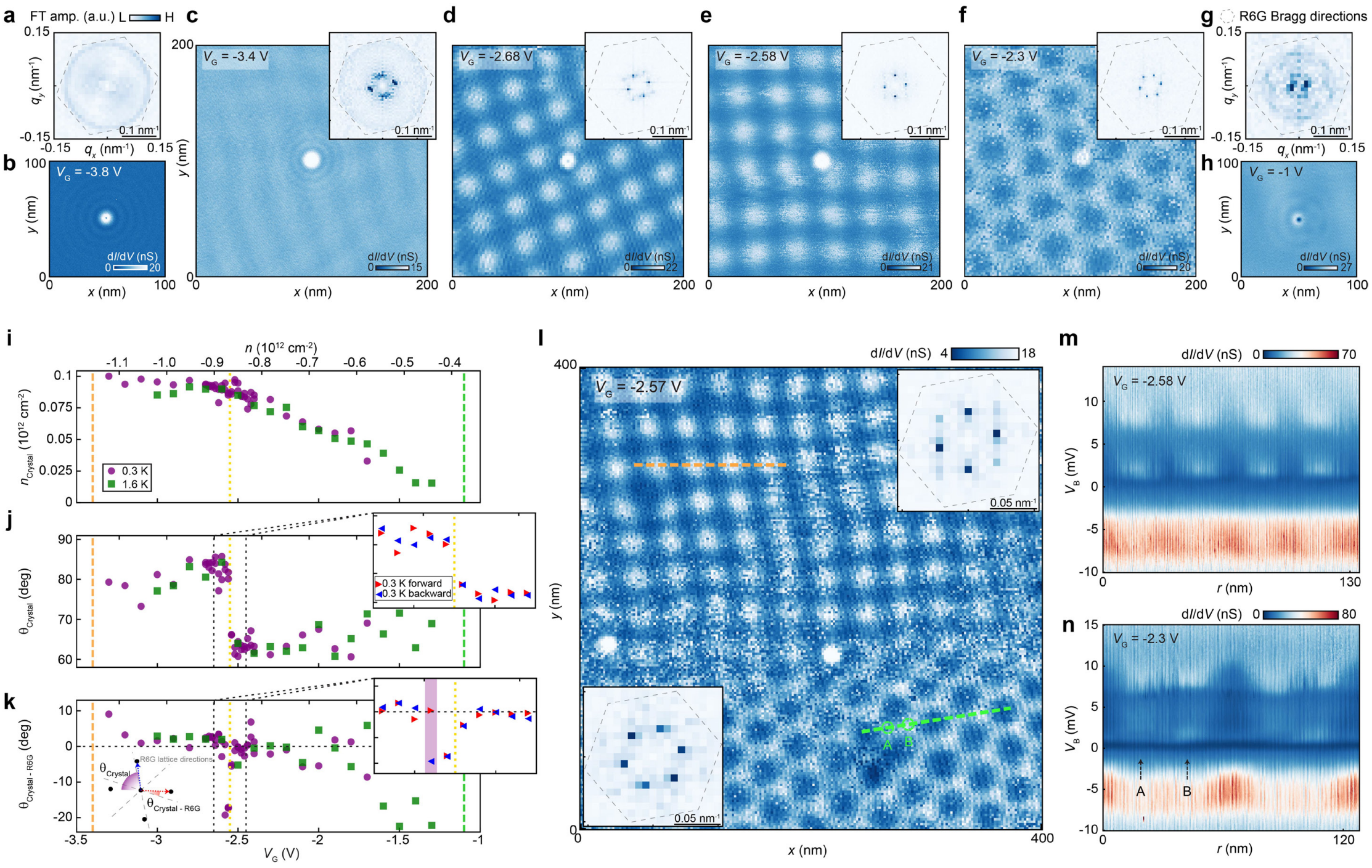

**Fig. 2 | STS visualization of oblique and honeycomb electron crystals. a-h,** Representative d$I$/d$V$ images within (**c-f**) and outside (**a**, **b**, **g**, **h**) the electron crystal phase boundaries with corresponding Fourier transforms. Grey dashed hexagons indicate the R6G Bragg vector directions, with their vertices pointing along those directions. Gaussian masks are applied to the Fourier transformation to reduce low frequency background (see Methods for details). For the color scale, H denotes high; L denotes low. **i-k**, Crystal density $n_{\mathrm{Crystal}}$, inter-vector angle $\theta_{\mathrm{Crystal}}$ and misalignment angle $\theta_{\mathrm{Crystal-R6G}}$ relative to the R6G lattice as functions of gate voltage at 0.3 K (purple

circles) and 1.6 K (green squares). Crystal phase boundaries are marked by color-coded dashed lines corresponding to those in Fig. 1f. Insets in **j**, **k** show zoom-ins near the oblique-to-honeycomb transition for forward (red triangles) and backward (blue triangles) gate-voltage sweeps, revealing hysteresis in $\theta_{\mathrm{Crystal-R6G}}$. Left inset in **k** illustrates the definitions of $\theta_{\mathrm{Crystal}}$ and $\theta_{\mathrm{Crystal-R6G}}$, which are described in detail in the Methods. **l,** d$I$/d$V$ map showing coexisting oblique and honeycomb domains at the oblique-to-honeycomb transition. The upper and lower insets show the Fourier transforms of the corresponding upper and lower domains, respectively. **m**, **n**, Spectroscopic linecuts across the high symmetry points of the oblique (**m**) and honeycomb (**n**) lattices at 0.3 K. The linecut directions are schematically indicated in **l** by the orange and green dashed lines, respectively. The honeycomb lattice displays a pronounced asymmetry in the spectra between the two sublattices, labeled A and B. d$I$/d$V$ measurement bias voltages: $V_{\mathrm{B}}$ = 0 mV in **a-c** and **g-h**; $V_{\mathrm{B}}$ = 2 mV in **d-f** and **l**. Tip height setpoints: $V_{\mathrm{B}}$ = 100 mV, $I_{\mathrm{t}}$ = 100 pA, $\Delta z$ = -160 pm in **a-c** and **g-h**; $V_{\mathrm{B}}$ = 100 mV, $I_{\mathrm{t}}$ = 5 nA in **d-f** and **l;** $V_{\mathrm{B}}$ = 100 mV, $I_{\mathrm{t}}$ = 5 nA in **m-n**.

**Melting of the electron crystal and temperature–density phase diagram**

We next examine the thermal melting of the two crystal phases and map out their temperature–density phase diagram. Figure 3a–c show current images of the oblique lattice at $V_{\mathrm{G}} = -2.7$ V for $T$ = 0.3, 2.5, and 3.3 K. At the lowest temperature, a highly ordered crystal extends across the entire 400-nm-wide field of view. Upon warming to 2.5 K, the lattice contrast decreases, and the spatial pattern becomes less uniform: the crystal is most pronounced near the top and lower-left regions of the image, while the intervening area develops stripe-like features. At 3.3 K, slightly above the transition temperature, the crystal melts into a liquid, leaving only short-range remnants of oblique order near impurities. Figure 3d shows the corresponding evolution of the current profile along a representative linecut, revealing a continuous reduction in crystal contrast with increasing temperature. To quantify the melting, we Fourier transform the real-space images at various temperatures (Extended Data Fig. 9) and extract the amplitudes of the crystal Bragg peaks, shown in Fig. 3i, which quantify the relative current modulation. For both reciprocal-lattice vectors $q_1$ and $q_2$ (red and blue), the peak amplitudes decrease monotonically with temperature and drop sharply near the critical temperature of 3.1 K. Notably, the $q_2$ peak remains consistently stronger than the $q_1$ peak over the intermediate-temperature range, reflecting the emergence of stripe-like correlations during melting.

The honeycomb phase exhibits markedly different temperature dependence. At the lowest temperature, $T$ = 0.3 K, the large-area image contains domain- and stripe-like structures (Fig. 3g). As the temperature approaches the melting transition, the honeycomb crystal becomes increasingly ordered, developing a clear lattice pattern at $T$ = 3.8 K across the entire 400-nm-wide field of view (Fig. 3f). Upon further warming to 4.2 K (Fig. 3e), the long-range order rapidly disappears, leaving only localized remnant correlations near impurities. The nonmonotonic temperature dependence is further illustrated by the linecuts shown in Fig. 3h and by the Bragg peak amplitudes of the crystal extracted from a series of two-dimensional maps (Fig. 3j, Extended Data Fig. 9). The three peaks, $q_1$, $q_2$, and $q_3$, exhibit different behaviors at low temperatures due to the stripe-like structures, but converge near the transition as the crystalline order becomes more uniform and pronounced. Upon further warming, all three peak amplitudes collapse as the crystal melts at 4.2 K. This unusual behavior—crystalline order strengthening, rather than weakening, upon heating—is reminiscent of the Pomeranchuk effect, in which entropy associated with internal degrees of freedom stabilizes the solid phase over a degenerate Fermi liquid at elevated temperatures[19,48–51].

We further map the phase diagram of the two crystal phases using a series of two-dimensional maps acquired across gate voltages at $T$ = 0.3, 1.6, 3, 4 and 4.14 K (Extended Data Fig. 6, 8, 9). In Fig. 3k, the onset of the oblique and honeycomb phases and the oblique-to-honeycomb transition are marked by orange squares, green hexagons, and yellow circles, respectively. The critical carrier density for the oblique-to-honeycomb transition is nearly temperature independent, whereas the density ranges of both crystal phases narrow with increasing temperature. The oblique lattice appears more fragile than the honeycomb lattice upon warming. Within the crystal phase boundaries, the intensity of the orange and green shading schematically represents the measured relative current-modulation amplitude (Extended Data Fig. 10). The Pomeranchuk-like behavior is most pronounced in the low-doping regime ($V_G \approx -1.6$ to $-1.1$ V). As illustrated by the images acquired at $V_G = -1.6$ V in Extended Data Fig. 10, structural fluctuations obscure nearly all periodic order at 0.3 K, whereas a robust crystalline pattern emerges upon warming to 1.6 K.

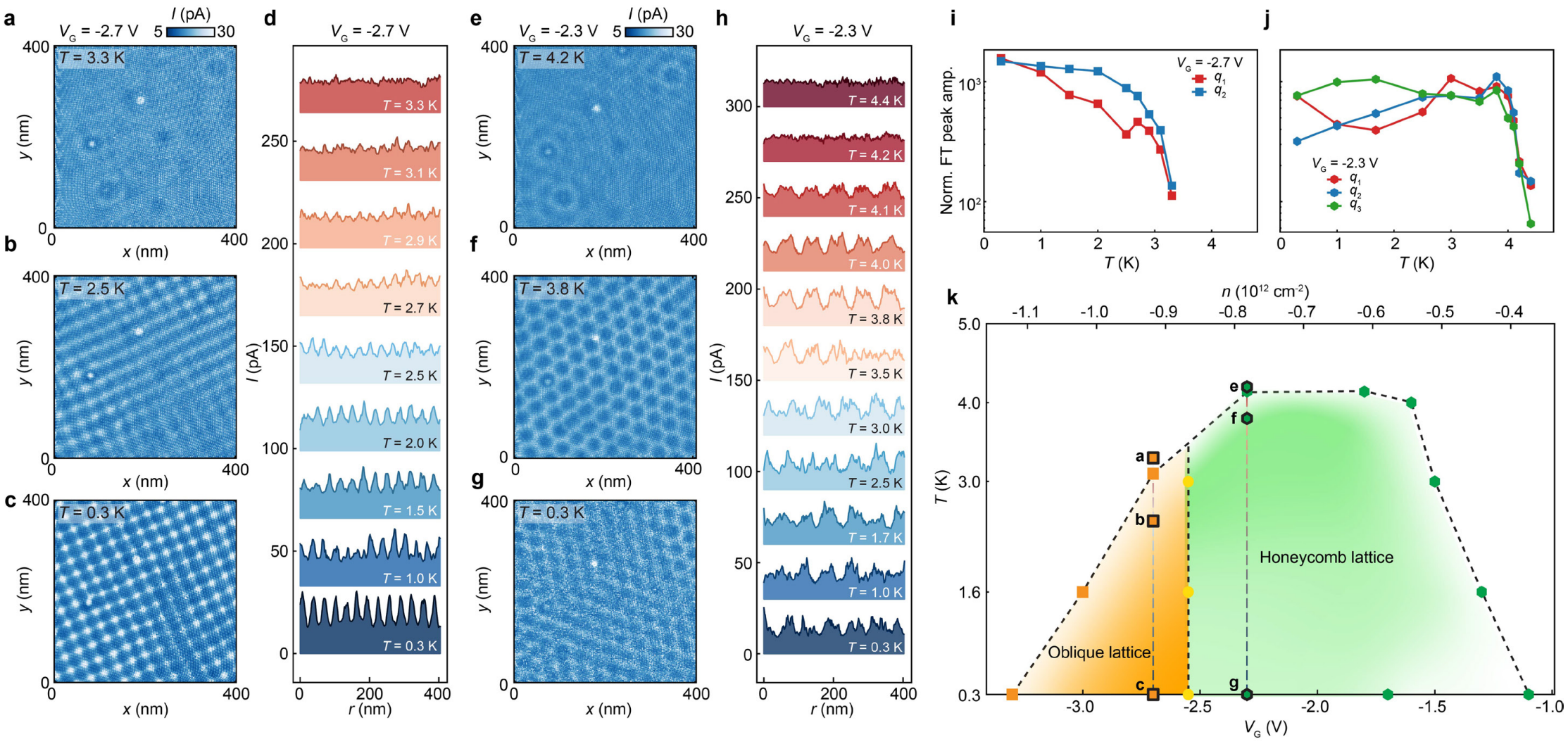

**Fig. 3 | Thermal melting of electron crystals and the temperature-density phase diagram. a–c**, **e–g**, Real-space tunneling current images acquired at $V_B$ = 2 mV for the oblique crystal (**a–c**, $V_G$ = −2.7 V) and the honeycomb crystal (**e–g**, $V_G$ = −2.3 V) at different temperatures. Setpoint parameters: $V_B$ = 100 mV, $I_t$ = 1 nA. **d**, **h**, Representative linecuts through high-symmetry points of the crystal lattices at various temperatures. The linecuts are smoothed using a Savitzky–Golay filter. **i**, **j**, Normalized Fourier peak amplitudes of the crystal lattices (see Methods) as a function of temperature, revealing the evolution of crystal order strength with changing temperatures. **k**, Temperature-gate-voltage phase diagram of the electron crystals. The color intensity of the orange and green shading schematically represents the modulation

strength of the crystal pattern. The experimentally determined boundaries of the oblique and honeycomb phases are marked by orange squares and green hexagons, respectively, while the oblique-to-honeycomb transition points are marked by yellow circles. The gate-voltage and temperature conditions corresponding to **a**–**c** and **e**–**g** are indicated by black-outlined orange squares and green hexagons, respectively. The complete source dataset is shown in Extended Data Fig. 6, 8, 9.

**Magnetic properties of the electron crystals**

The response of the oblique and honeycomb crystals to a perpendicular magnetic field suggests that the spatial symmetry breaking in these phases may be intertwined with time-reversal-symmetry-breaking orbital magnetism. Figure 4a, b compare the oblique crystal at $V_G = -2.7$ V under zero field and at $B = 1.5$ T. At 1.5 T, a clear $\sqrt{2} \times \sqrt{2}$ superstructure emerges in the real-space image, which is also evident in its Fourier transform (Fig. 4b, inset). Within the doubled real-space unit cell (marked by the orange square in Fig. 4b), the two crystal sublattices, A and B, which are equivalent at $B = 0$ T, develop an intensity contrast. Spectroscopic measurements across the two sublattices reveal a field-induced difference in their spectral responses (Fig. 4c, d). In particular, the near-zero-energy peaks that underpin the crystal pattern (solid and dashed lines in Fig. 4e; see also Extended Data Fig. 11) split between the two sublattices by an energy $\Delta_{AB}$. The splitting increases linearly as the magnetic field increases (Fig. 4f), corresponding to an effective $g$-factor of 16.0±1.1. This large $g$-factor, far exceeding the free-electron value, indicates that the splitting is unlikely to arise from spin Zeeman coupling and is instead consistent with opposite orbital magnetization on the two crystal sublattices. Upon further increasing the field to $B = 2.5$ T, the electron crystal melts into a disordered state (Extended Data Fig. 11). These observations suggest that the oblique crystal possesses an alternating orbital magnetization in addition to charge order.

The honeycomb crystal responds differently to a perpendicular magnetic field: a field as small as $B = 50$ mT noticeably sharpens the honeycomb pattern (Fig. 4g, h, Extended Data Fig. 12). Moreover, the honeycomb crystal emerges in the same region of the phase diagram where transport measurements previously identified orbital magnetism (Extended Data Fig. 2)[10]. The marked response to a small magnetic field may be interpreted as field-induced alignment of orbital magnetic domains. Applying a larger magnetic field has little further effect on the spatial pattern (Extended Data Fig. 12c), consistent with the orbital magnetization being unipolar in real space.

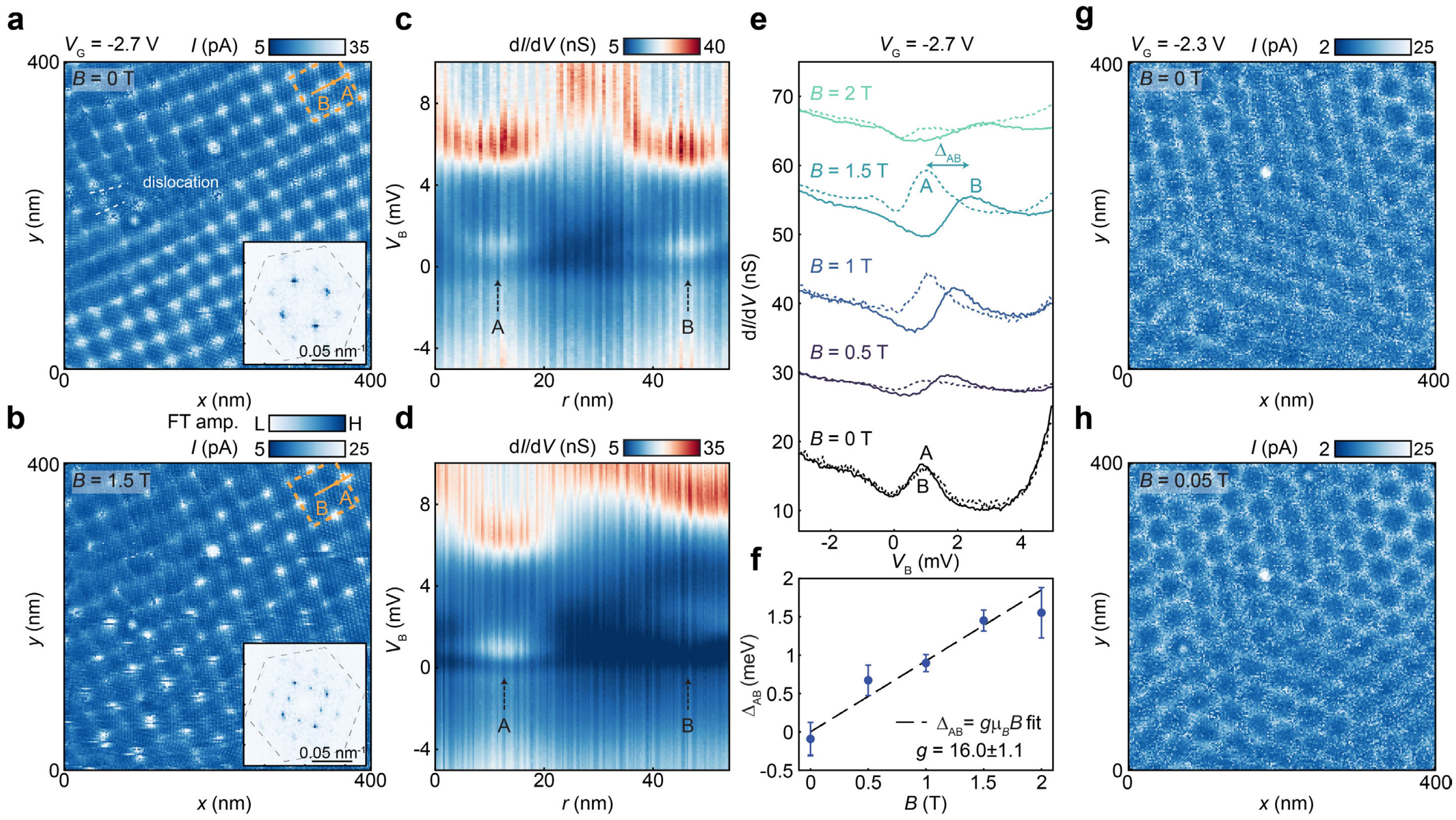

**Fig. 4 | Magnetic responses of the electron crystals. a**, **b**, Real-space tunneling current images of the oblique crystal acquired at $V_G = -2.7$ V and $V_B = 2$ mV under $B = 0$ T and $B = 1.5$ T. For $B = 1.5$ T, the crystal unit cell is enlarged by a factor of $\sqrt{2}$ and doubles in area. The enlarged unit cell, outlined by an orange dashed square, contains two sublattice sites labeled A and B. Insets show corresponding Fourier transforms. An edge dislocation is present toward the left side of these images, locally reducing the contrast of the surrounding crystal modulation. H: high; L: low. Setpoint parameters: $V_B = 100$ mV, $I_t = 1$ nA. **c**, **d**, Spectroscopic linecuts through the two sublattice sites, A and B, of the enlarged unit cell. The linecut locations are indicated by orange arrows in **a** and **b**. **e**, Tunneling spectra acquired at site A (dashed curves) and site B (solid curves) at various magnetic fields. The peak energy splitting between the two sublattices is denoted by $\Delta_{AB}$. **f**, Magnetic field dependence of the peak energy splitting $\Delta_{AB}$. A linear fit to $\Delta_{AB} = g\mu_B B$ yields $g = 16.0 \pm 1.1$. Error bars represent the standard deviations of the extracted peak energies. **g**, **h**, Real-space tunneling current images of the honeycomb lattice at $V_G = -2.3$ V and $V_B = 2$ mV under $B = 0$ T and $B = 0.05$ T, revealing a noticeable suppression of fluctuations under a field of only 50 mT. Setpoint parameters: $V_B = 100$ mV, $I_t = 1$ nA.

## Discussion and outlook

We propose a phenomenological picture that may account for our experimental observations. For the crystals observed here, the majority of carriers remain itinerant, whereas a minority crystallize. In the honeycomb phase, the crystallized carriers may take the form of an anomalous Hall crystal with spontaneous valley polarization. Owing to the valley-contrasting Berry curvature, this state possesses a spontaneous orbital magnetic moment, whose polarity can be selected by a small perpendicular magnetic field, and produces an anomalous Hall response. The coexistence with itinerant carriers provides a natural explanation for why the anomalous Hall response is not quantized. The oblique phase, in contrast, could be interpreted as an orbital-antiferromagnetic crystal. In this state, the orbital magnetic moments alternate between neighboring sites of the spontaneously formed lattice, analogous to Néel order but involving orbital rather than spin magnetization. We refer to this state as an orbital-antiferromagnetic crystal. The cancellation of these opposing orbital moments naturally accounts for the absence of an anomalous Hall response.

Our observations reveal a new paradigm of electron crystallization and raise the possibility that electron crystallization can be intertwined with orbital magnetism and topological phenomena. Further studies are needed to fully establish this connection and determine whether the real-space crystalline order underlies the emergence of fractional Chern insulators and candidate chiral superconductivity in rhombohedral graphene systems.

## Methods

### Sample fabrication

The rhombohedral hexalayer graphene devices were prepared with methods similar to previous literature[10,52–54]. Supporting structures, including contact graphite, hBN, and bottom graphite gate, were sequentially picked up using a polyvinyl alcohol (PVA) stamp and released onto $SiO_2$-Si prepatterned substrates with Au/Cr electrodes. They were then cleaned with HPLC water, acetone, isopropyl alcohol (IPA), and *n*-methyl-2-pyrrolidone and annealed in forming gas (hydrogen/nitrogen) at around 350 °C for more than 10 hours. R6G flakes were identified through optical contrast (for layer number) and Raman spectroscopy (for rhombohedral stacking), and then isolated through anodic oxidation lithography using an atomic force microscope[55]. We picked up the R6G with another PVA stamp, released it on the annealed supporting structure, and washed it again with HPLC water and IPA. The final device was annealed in forming gas and then in an ultrahigh vacuum chamber at around 350 °C for more than 10 hours before STM measurements. We monitored the rhombohedral stacking order throughout the fabrication process (Extended Data Fig. 13).

### STM/STS measurements

All STM/STS measurements were performed in a Unisoku 1300 STM system with a base temperature of approximately 300 mK. We used freshly electrochemically etched tungsten tips for this study. The tips were first conditioned by field emission, followed by controlled indentation on a clean Cu(111) surface until the poke mark was confined and an expected spectrum showing Cu(111) surface state was obtained. For all tunneling measurements, a 10:1 divider was applied to the bias line, which attenuates the output voltage by a factor of 10 and improves the energy resolution. Bias voltages reported in this work are the values after the divider correction. Tunneling spectra were acquired using a standard lock-in technique with reference frequencies $f_{\mathrm{ref}}$ = 907–932 Hz and modulation amplitudes $V_{\mathrm{rms}}$ = 0.4–1 mV.

To obtain spectroscopic images, including the QPI and electron crystal patterns, we first set the tip height at each $x$ and $y$ position with a setpoint voltage $V_{\mathrm{B}}$ and setpoint current $I_{\mathrm{t}}$. We then tune off the feedback to lock the tip height and acquire a spectrum before moving to the next point and repeating. For measurements with an indicated $\Delta z$, after disabling the feedback loop, we lower the tip height by $\Delta z$ before acquiring the tunnelling spectrum. This way, we could use a large setpoint voltage of $V_{\mathrm{B}}$ = 100 to 1000 mV to set the tip-sample distance while achieving a large signal for the spectral measurement. For electron crystal imaging, a high setpoint voltage makes sure the tip height tracks the sample height variations and reduces setpoint effect. Simultaneously acquired topographic images showed no discernible electron-crystal contrast, confirming setpoint effect is minimized.

After obtaining the position-dependent spectra, we show the electron-crystal structures with either the differential conductance d$I$/d$V$ or tunneling current $I_{\mathrm{t}}$ at a particular bias voltage $V_{\mathrm{B}}$, as specified in the corresponding figure captions. The two channels exhibit qualitatively similar spatial patterns at a low positive sample bias of $V_{\mathrm{B}}$ = 2 mV.

### Corbino transport measurement with the tip

After finishing the tunneling measurements, we perform *in situ* Corbino transport measurements by lowering the tip so it becomes an ohmic contact[53]. To do this in a controlled way, the STM tip was first stabilized in the tunneling regime at a setpoint of $V_B = 1$ V and $I_t = 100$ pA. The tip was held under these conditions for at least 30 minutes to minimize piezoelectric drift. Once the drift became negligible, the feedback loop was disabled, and the tip was manually lowered toward the sample by $\Delta z = 2$–$3$ nm at $V_B = 0$ V to establish electrical contact. The Corbino conductance was measured using a standard lock-in technique with a reference frequency of 932 Hz and a modulation voltage $V_{rms} = 0.4$ mV. In the highly doped regime at zero magnetic field, this excitation generated a current of approximately 30–40 nA, corresponding to an effective tip-sample resistance of approximately 10 kΩ, consistent with a highly transmitting point-contact junction.

### Fourier transform and Gaussian filtering

The Fourier transform of real-space d$I$/d$V$ or current images is numerically calculated using the Fast Fourier Transform (FFT) algorithm. A Gaussian mask centered at zero frequency is sometimes applied in Fourier space to improve the contrast of the $q$-space spectra (e.g., Fig. 2l, 4a, b). For QPI measurements (Fig. 1b–d, 2a–g; Extended Data Fig. 3c–f), a separate real-space Gaussian mask was applied around the defect before the FFT to exclude the strong non-periodic spectral response of the defect itself.

### Extraction of electron crystal parameters

We extract the reciprocal lattice vectors from the peaks in the Fourier transforms of the crystal patterns. For both oblique and honeycomb lattices, we select the two prominent conjugate reciprocal vectors, $\mathbf{q}_1$ and $\mathbf{q}_2$. Then we obtain the real-space electron crystal lattice vectors $\mathbf{a}_i$ from the reciprocal lattice vectors: the corresponding real-space primitive vectors are given by the reciprocal relation: $\begin{pmatrix}\mathbf{a}_1\\ \mathbf{a}_2\end{pmatrix} = 2\pi\left[\begin{pmatrix}\mathbf{q}_1\\ \mathbf{q}_2\end{pmatrix}^{-1}\right]^T$. Compared to directly reading out $\mathbf{a}_i$ from real space images, this $q$-space approach allows us to average out local distortions and fluctuations[45]. For honeycomb lattices, a third vector $\mathbf{a}_3$ is defined as one of the linear combinations $\{\mathbf{a}_1+\mathbf{a}_2, \mathbf{a}_1-\mathbf{a}_2\}$. A similar method is used to obtain the R6G lattice vectors and reciprocal vectors.

The crystal unit cell density (Fig. 2i) is calculated as $n_{\text{Crystal}} = \frac{1}{|\mathbf{a}_1\times\mathbf{a}_2|}$, representing the inverse of the cell area. The internal crystal angle $\theta_{\text{Crystal}}$ (Fig. 2j), which is the angle between the two lattice vectors, is computed as $\theta_{\text{Crystal}} = \arccos\left(\frac{|\mathbf{a}_1\cdot\mathbf{a}_2|}{a_1 a_2}\right)$. To determine the orientation relative to the R6G lattice, $\theta_{\text{Crystal–R6G}}$ (Fig. 2k), we designate the real-space R6G vector closest to the $x$-axis as the reference vector, $\mathbf{a}_{\text{R6G}}$. The relative orientation is then defined as the minimum-magnitude signed angle between the set $\{\mathbf{a}_1, \mathbf{a}_2, \mathbf{a}_3\}$ and $\mathbf{a}_{\text{R6G}}$.

### Fourier amplitude of crystal Bragg peaks

The crystal Bragg peak strengths shown in Fig. 3i, j are extracted from the Fourier transforms of the crystal patterns. We note that spectral leakage across multiple pixels, together with the mismatch between the discrete Fourier sampling grid and the true Bragg-point positions, can cause inaccuracy in the measured Bragg peak amplitudes, $A(\mathbf{q}_i)$. To minimize this, we approximate the true amplitude by calculating the root-mean-square over an empirical local neighborhood $\Omega_{\mathbf{q}_i}$ (spanning several pixels, as also used in previous studies[56,57]) centered on $\mathbf{q}_i$: $A^{\text{approx}}(\mathbf{q}_i) = \sqrt{\frac{1}{|\Omega_{\mathbf{q}_i}|}\sum_{\mathbf{q}\in\Omega_{\mathbf{q}_i}}|A(\mathbf{q})|^2}$. This approach leverages Parseval's theorem to conserve the localized signal power irrespective of grid mismatches. We then normalize the extracted amplitudes by the average current of each image. This normalization accounts for variations in the overall signal strength arising from different tunneling setpoints and measurement conditions, enabling more reliable comparison of data acquired at the same gate voltage.

## Acknowledgements

We acknowledge helpful discussions with Long Ju, Jia Li, Leonid S. Levitov, Taige Wang, Filippo Gaggioli, Daniele Guerci, Ahmed Abouelkomsan, and Liang Fu.

X.L. and Y.S. acknowledge support from the National Science Foundation through CAREER Award No. DMR-2442363. J.D. acknowledges support from the New Frontier Grant, College of Arts and Sciences, Cornell University. K.W. and T.T. acknowledge support from the CREST (JPMJCR24A5), JST and World Premier International Research Center Initiative (WPI), MEXT, Japan. This work was performed in part at the Cornell NanoScale Facility, a member of the National Nanotechnology Coordinated Infrastructure (NNCI), which is supported by the National Science Foundation (Grant NNCI-2025233). This work also made use of the Cornell Center for Materials Research shared instrumentation facility.

## Author contributions

X.L. and J.D. conceived the project. Y.S., J.X., and D.G. fabricated the device. Y.S. and J.D. performed the measurements, analyzed the data, and prepared the figures with input from X.L. K.W. and T.T. provided hBN crystals. Y.S., J.D., and X.L. wrote the manuscript with input from all authors.

## Competing interests

The authors declare no competing interests.

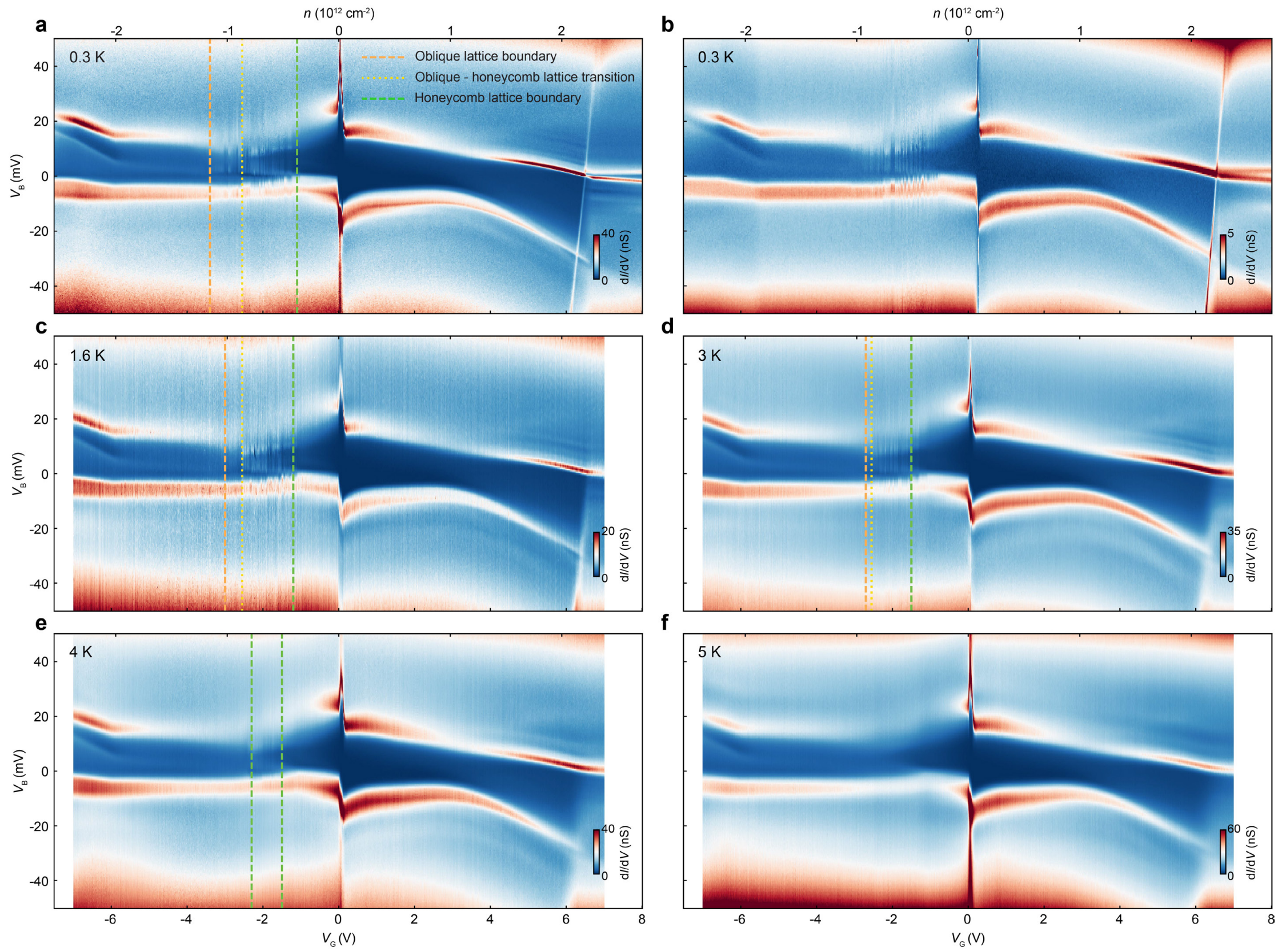

**Extended Data Fig. 1 | Reproducibility and temperature dependence of gate-dependent point-tunneling spectra d$I$/d$V$($V_G$, $V_B$). a**, **b**, Two sets of d$I$/d$V$($V_G$, $V_B$) data obtained with different tips and on different sample locations at 0.3 K. Both datasets reproduce the same overall spectroscopic phase diagram, including the insulating state near zero gate voltage and pronounced spectroscopic fluctuations over a similar gate-voltage range that coincides with the electron crystal phase. Setpoint and measurement conditions for **a**: $V_B$ = 100 mV, $I_t$ = 1 nA, $\Delta z$ = -20 pm, lock-in modulation voltage $V_{rms}$ = 0.5 mV. For **b**: $V_B$ = 1 V, $I_t$ = 100 pA, $\Delta z$ = -350 pm, $V_{rms}$ = 1 mV. **c**, 1.6 K d$I$/d$V$($V_G$, $V_B$) data with constant tip height (reference setpoint: gate voltage $V_G$ = -7 V, $V_B$ = 100 mV, $I_t$ = 2 nA, $V_{rms}$ = 1 mV, see our previous study for methods[58]). **d**, 3 K d$I$/d$V$($V_G$, $V_B$) data with setpoint $V_B$ = 100 mV, $I_t$ = 3 nA, $V_{rms}$ = 1 mV. **e**, 4 K d$I$/d$V$($V_G$, $V_B$) data with setpoint $V_B$ = 100 mV, $I_t$ = 3 nA, $V_{rms}$ = 1 mV. **f**, 5 K d$I$/d$V$($V_G$, $V_B$) data with setpoint $V_B$ = 100 mV, $I_t$ = 4 nA, $\Delta z$ = -20 pm, $V_{rms}$ = 1 mV. Phase boundaries of oblique and honeycomb lattices extracted from STS images are marked by dashed lines and coincide with spectroscopic fluctuations (verified in multiple rounds of measurements such as **a**, **b**). With increasing temperature, the fluctuation region and crystal phases shrink simultaneously. Note that the oblique lattice disappears first with rising temperature, thus only the honeycomb crystal persists at 4 K (**e**, also see Fig. 3).

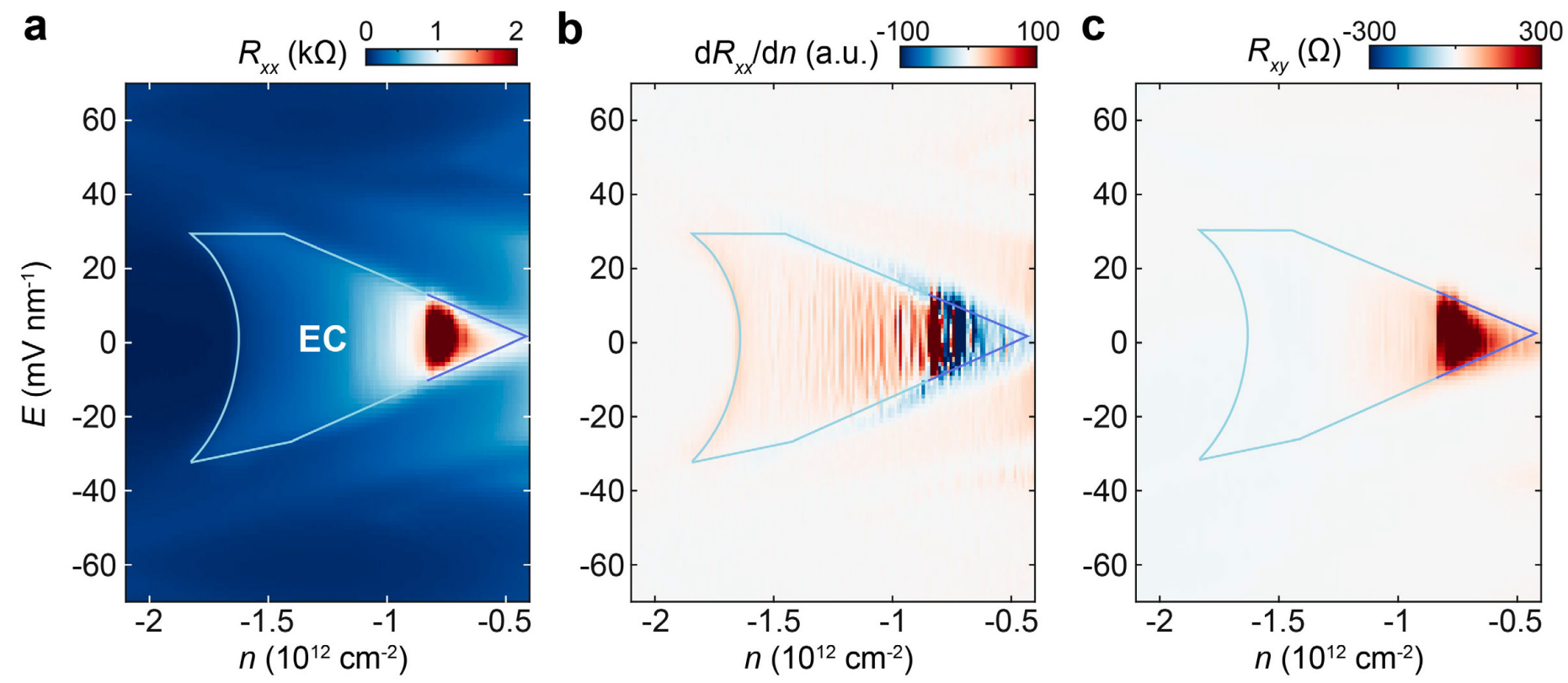

**Extended Data Fig. 2 | Transport signatures of resistance fluctuations and anomalous Hall response in a separate dual-gated R6G device in the electron crystal regime. a,** Longitudinal resistance, $R_{xx}$, as a function of carrier density $n$ and electric field $E$, measured in a dual-gated Hall-bar device at a base temperature of 10 mK and under $B = 50$ mT. The electron crystal (EC) regime is indicated by the outlines, with light and dark blue highlighting two phases without and with an anomalous Hall response that align with STM observations. A detailed characterization of this device is reported in our previous study[10]. **b,** Numerical derivative $dR_{xx}/dn$ of the data shown in **a**, highlighting resistance fluctuations within the triangular regions associated with the electron crystal phases. **c,** Hall resistance, $R_{xy}$, as a function of $n$ and $E$. A pronounced anomalous Hall response is present in the low-density section of the triangular region, where $|n| \lesssim 0.8 \times 10^{12}$ cm$^{-2}$. We note that the portion of the electron-crystal triangle exhibiting an anomalous Hall response varies from device to device, with some devices showing a substantially larger anomalous Hall phase space within the electron crystal triangle[59].

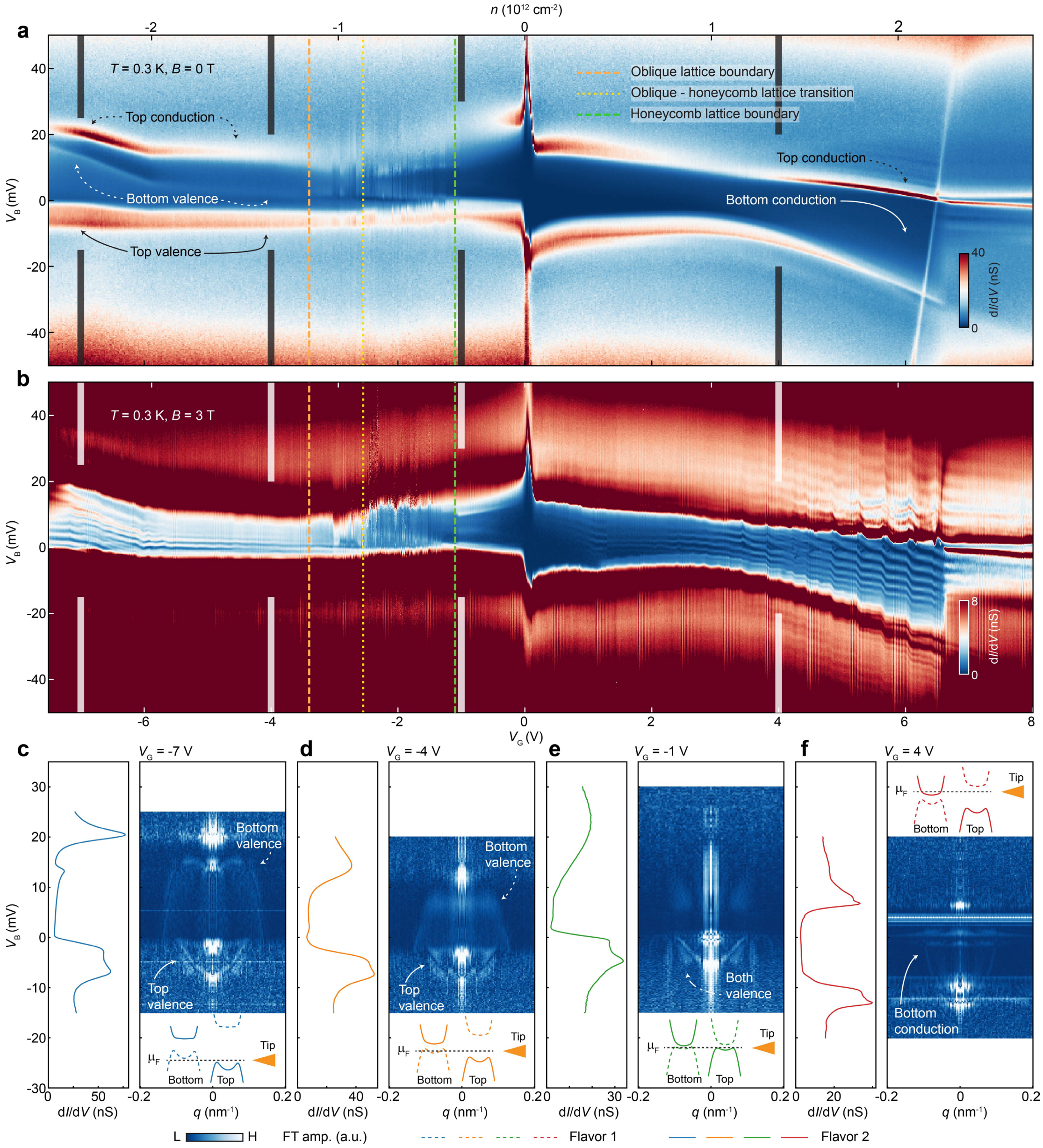

**Extended Data Fig. 3 | Analysis of the R6G band structure with gate-dependent point spectra and quasiparticle interference (QPI) spectra. a**, **b**, d$I$/d$V$ as a function of gate and bias voltages at $B$ = 0 T (**a**) and 3 T (**b**). The boundaries of the oblique and honeycomb electron crystal phases are indicated by dashed lines. Setpoint conditions: $V_{\mathrm{B}}$ = 1 V, $I_{\mathrm{t}}$ = 100 pA, $\Delta z$ = −350 pm (**a**) and −400 pm (**b**); lock-in modulation voltage $V_{\mathrm{rms}}$ = 0.4 mV. **c-f**, QPI dispersion at representative gate voltages of $V_{\mathrm{G}}$ = −7 V (**c**), −4 V (**d**), −1 V (**e**) and 4 V (**f**). The QPI dispersions are obtained by Fourier transforming d$I$/d$V$ linecuts across a point defect with a real-space Gaussian mask applied around the defect center (see Methods for details). The d$I$/d$V$ spectra, spatially averaged over regions away from the defect center, are shown to the left of each QPI dispersion on the same bias axis for comparison. The corresponding gate voltage positions are marked in **a** and **b** with vertical black and white lines. Gaps in these lines indicate the bias ranges over which the QPI measurements are acquired. Setpoint parameters for **c, d**: $V_{\mathrm{B}}$ = 100 mV, $I_{\mathrm{t}}$ = 5 nA; $V_{\mathrm{rms}}$ = 0.2 mV. For **e**, $V_{\mathrm{B}}$ = 100 mV, $I_{\mathrm{t}}$ = 100 pA, $\Delta z$ = −165 pm; $V_{\mathrm{rms}}$ = 0.2 mV for hole excitations and $V_{\mathrm{B}}$ = 1 V, $I_{\mathrm{t}}$ = 100 pA, $\Delta z$ = −390 pm; $V_{\mathrm{rms}}$ = 0.4

mV for electron excitations. For **f**, $V_B$ = 100 mV, $I_t$ = 100 pA, $\Delta z = -170$ pm; $V_{rms}$ = 0.2 mV.

For rhombohedral graphene with $N > 3$ layers, interactions open a correlated gap between the Mexican-hat-shaped valence and conduction bands at zero electric field. Across the correlation gap, the valence and conduction bands are polarized toward opposite outer layers, with the layer polarization reversed between the two spin flavors. Increasing $|V_G|$ enlarges the gap for one spin flavor while reducing it for the other, as illustrated in **c-f**. Because the STM tip couples much more strongly to the top layer, spectral features associated with the top-layer bands appear substantially more pronounced than those associated with the bottom-layer bands. Meanwhile, the band crossing the Fermi energy is localized predominantly on the bottom layer at both positive and negative gate voltages. Consequently, the spectra exhibit an apparent gap-like feature around the Fermi energy throughout the gate-voltage range, bounded by strong signals from the top-layer conduction and valence bands.

On the hole-doped side ($V_G < 0$), the van Hove singularity of the bottom-layer valence band produces a visible trace within this apparent gap (marked by white dashed arrows in **a**). On the electron-doped side, the spectral response of the bottom-layer conduction band is too weak to be clearly resolved. Applying a magnetic field of $B$ = 3 T reveals the otherwise invisible bottom-layer conduction band features via the emergence of Landau levels within the apparently gapped region, confirming a finite density of states associated with these bands.

QPI dispersions corroborate the electric field evolution of top- and bottom-layer bands. In **e**, where the electric field is small, the valence band associated with the two flavors appears to overlap. In **c** and **d**, however, Mexican-hat-shaped valence bands of the two layers are energetically split. This splitting increases with increasing $|V_G|$ on the hole side, which is also evident in point spectra of **a**. The gate-induced electric field separates the valence bands on the two layers, leading to an increasing energy splitting between the top- and bottom-layer valence band van Hove singularities observed here. The $dI/dV$ spectra shown on the left exhibit enhanced density of states at the biases corresponding to the van Hove singularities identified from the QPI dispersion. Layer polarization (top or bottom) of band features is marked in **a** and **c**–**f**. It is also worth noting that the active sublattices of the top and bottom layers are located in the same *xy* positions for hexalayer rhombohedral graphene (Fig. 1a), causing the layer polarization not to be directly distinguished from atomic scale imaging (Extended Data Fig. 13).

Additionally, the point spectra show a transition near $V_G$ = -6 V (**a**, **b**), beyond which the trajectories of the bottom-band van Hove singularities become steeper. Comparing with the transport phase diagram suggests that this transition occurs when the Fermi level passes through the van Hove singularity of the bottom-layer band and enters a more dispersive portion of the band, resulting in the steeper gate dependence. Another abrupt transition occurs near $V_G$ = 6.5 V, where the Fermi level rises to reach the top-layer conduction band. Owing to the high density of states associated with the van Hove singularity near the band edge at higher $E$ field, the corresponding spectral feature becomes nearly independent of gate voltage. A small energy gap opens in the top-layer band at the Fermi energy, likely due to strong correlation. At the same time, the bottom conduction band still crosses the Fermi level, consistent with a partial isospin polarized state reported previously[10].

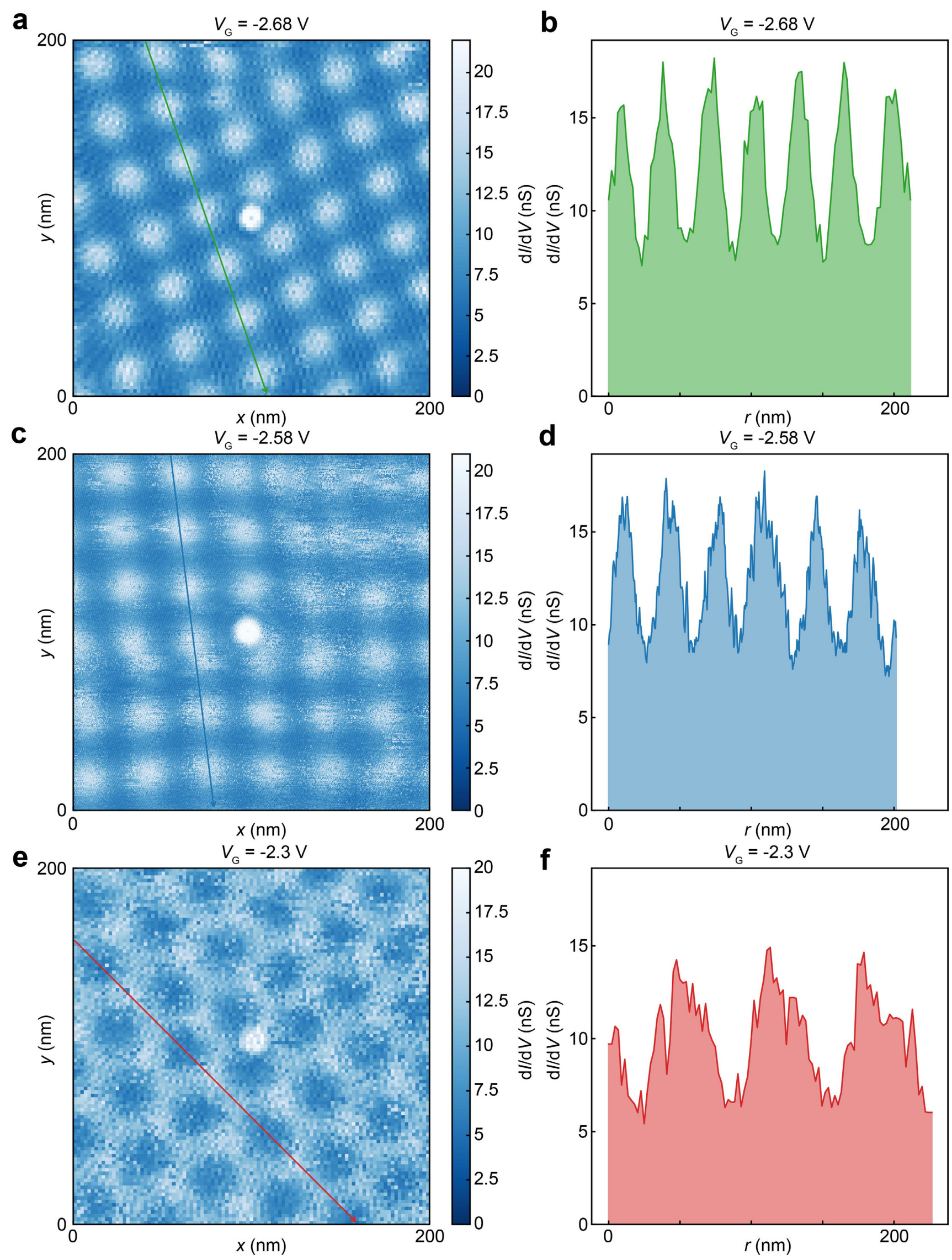

**Extended Data Fig. 4 | Linecuts showing the crystal modulation contrast at representative electron crystal states. a, c, e**, Real-space d$I$/d$V$ maps acquired at bias voltage $V_B$ = 2 mV and gate voltages $V_G$ = -2.68 V, -2.58 V, and -2.3 V, respectively. Setpoint parameters: $V_B$ = 100 mV, $I_t$ = 5 nA; $V_{rms}$ = 0.4 mV. **b, d, f**, Linecuts extracted along the trace marked in **a, c, e**. The modulation contrast, quantified by the ratio of the minimum to maximum d$I$/d$V$, is approximately 1:2. The $V_G$ = -2.58 V linecut is smoothed using a Savitzky–Golay filter.

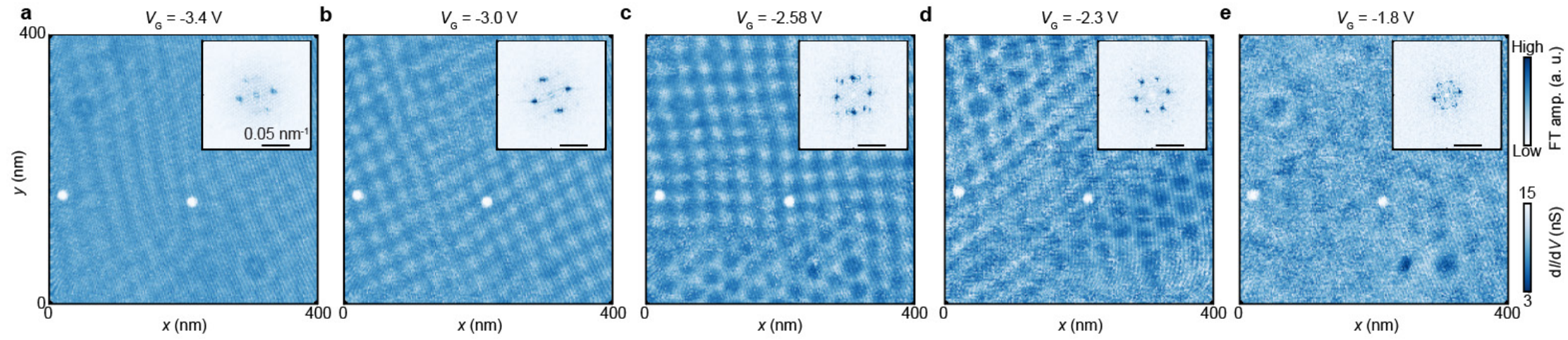

**Extended Data Fig. 5 | Representative large scale electron crystal images at various gate voltages. a-e,** Large scale d*I*/d*V* images obtained at bias voltage $V_B$ = 2 mV with setpoint $V_B$ = 100 mV, $I_t$ = 5 nA; $V_{rms}$ = 0.4 mV, at $V_G$ = -3.4, -3.0, -2.58, -2.3, -1.8 V, respectively. Insets show corresponding Fourier transform results.

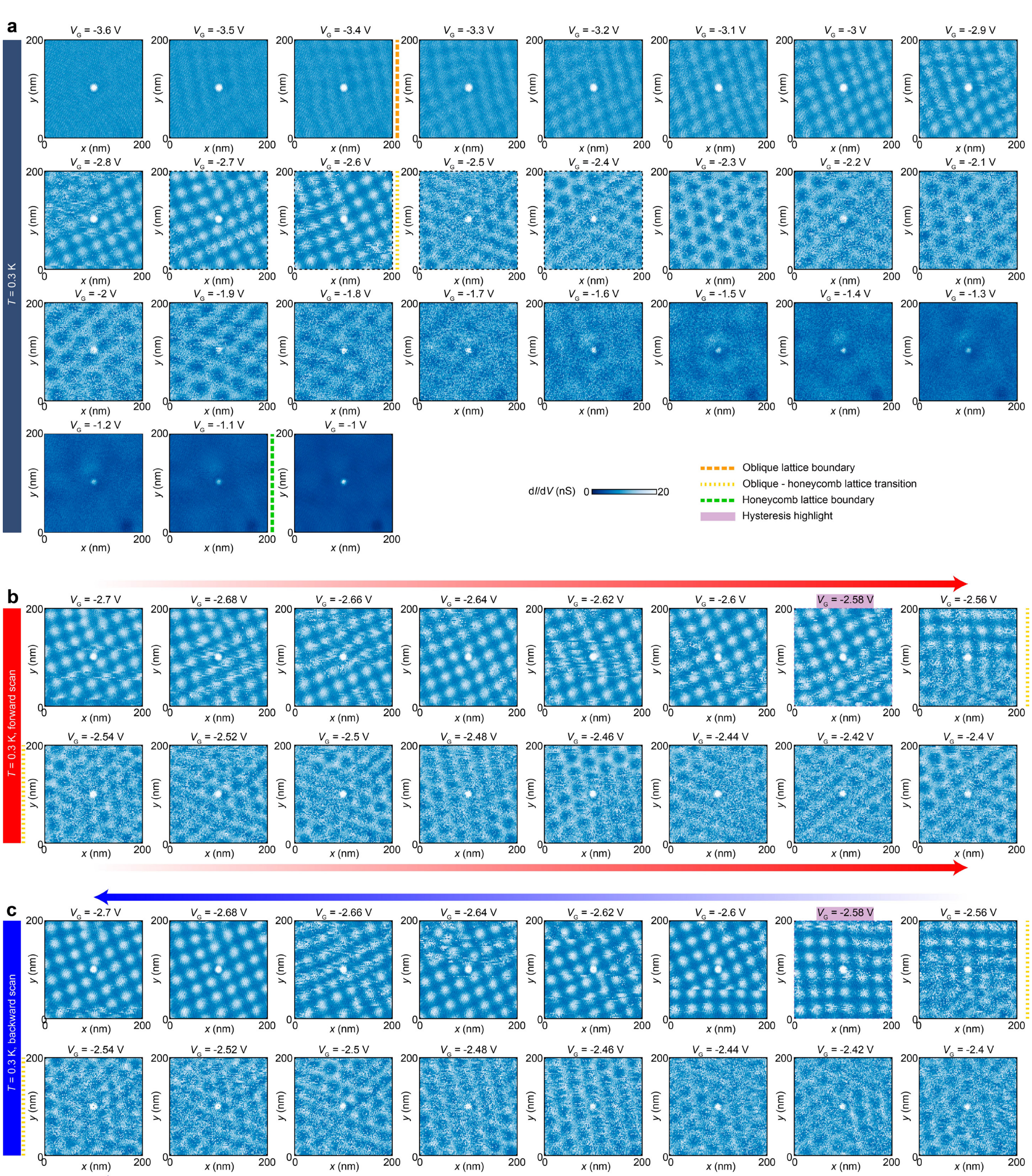

**Extended Data Fig. 6 | STS images across the electron crystal regime at 0.3 K. a,** d$I$/d$V$ maps acquired at various gate voltages with bias voltage $V_{\mathrm{B}}$ = 2 mV, $V_{\mathrm{rms}}$ = 0.4 mV, and setpoint $V_{\mathrm{B}}$ = 100 mV, $I_{\mathrm{t}}$ = 5 nA,. The phase boundaries are marked with dashed lines with corresponding colors. The zoomed-in gate-voltage range over which the hysteresis measurements in **b** and **c** are acquired is indicated by black dashed borders. **b, c,** STS images acquired during forward ($V_{\mathrm{G}}$ = -2.7 V to -2.4 V, **b**) and backward ($V_{\mathrm{G}}$ = -2.4 V to -2.7 V, **c**) gate-voltage sweeps, showing hysteresis in the crystal evolution around oblique-to-honeycomb transition at 0.3 K. The oblique-to-honeycomb transition is marked in both scans by yellow dashed lines. Hysteretic behavior associated with a change in the oblique crystal orientation is evident from highlighted images, which show different crystal orientations for the same gate voltage of $V_{\mathrm{G}}$ = -2.58 V. Arrows indicate direction of gate voltage sweeps. These d$I$/d$V$ maps are obtained at bias voltage $V_{\mathrm{B}}$ = 2 mV with setpoint $V_{\mathrm{B}}$ = 100 mV, $I_{\mathrm{t}}$ = 5 nA, $V_{\mathrm{rms}}$ = 0.4 mV.

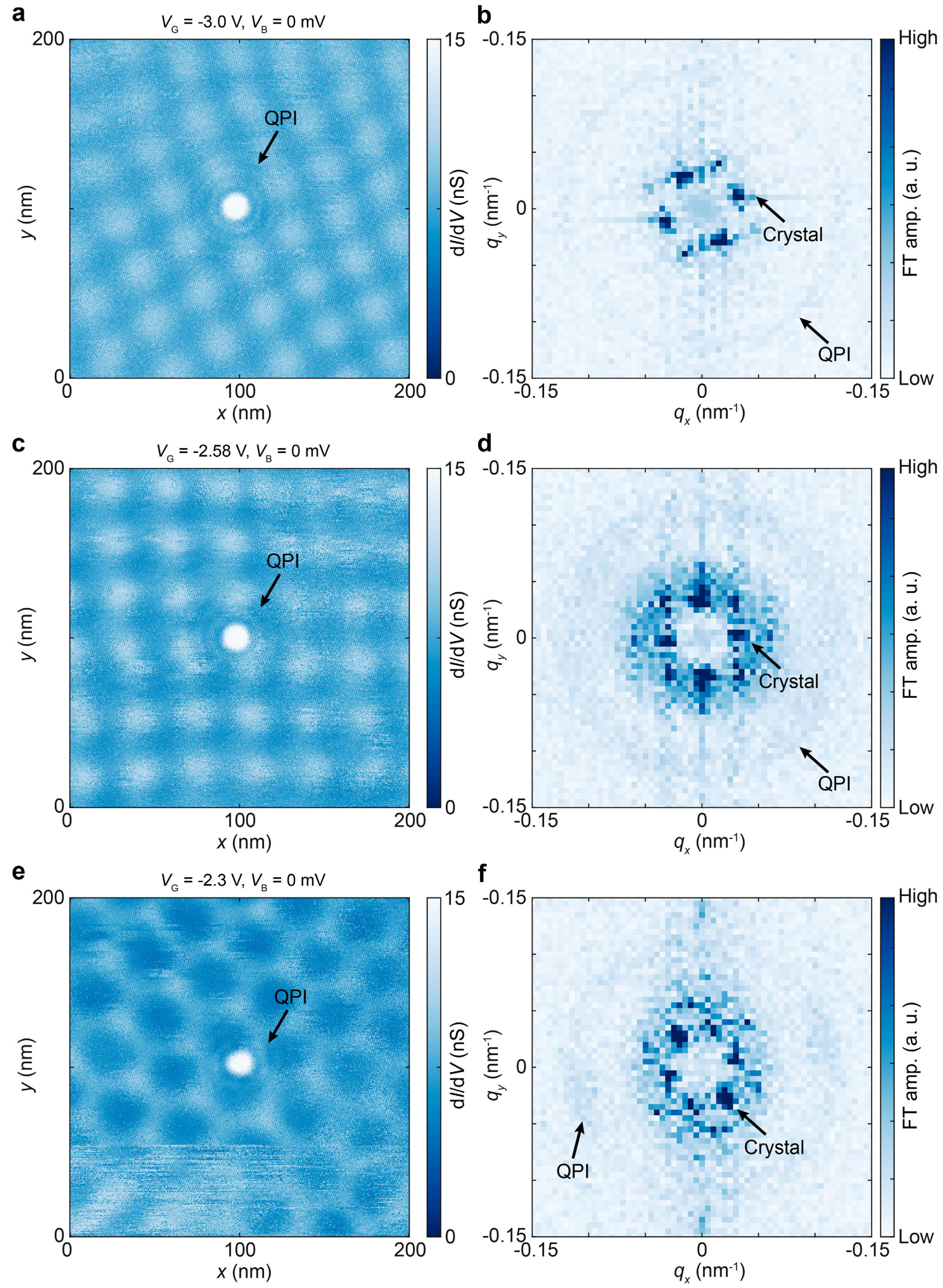

**Extended Data Fig. 7 | Coexistence of QPI and electron crystal order at the Fermi energy across different gate voltages. a-f,** d$I$/d$V$ images and corresponding Fourier transforms acquired at $V_G$ = -3 V (**a**, **b**), -2.58 V (**c**, **d**), -2.3 V (**e**, **f**), with $V_B$ = 0 mV at base temperature. Setpoint conditions: $V_B$ = 100 mV, $I_t$ = 5 nA; $V_{rms}$ = 0.4 mV. The QPI features are localized around the defect center in real space and give rise to ring-like structures in corresponding Fourier transforms, as indicated by the arrows. Peaks associated with the electron-crystal modulation coexist with these QPI features in the Fourier space. The QPI wavelengths appear much smaller than the crystal lattice vector lengths.

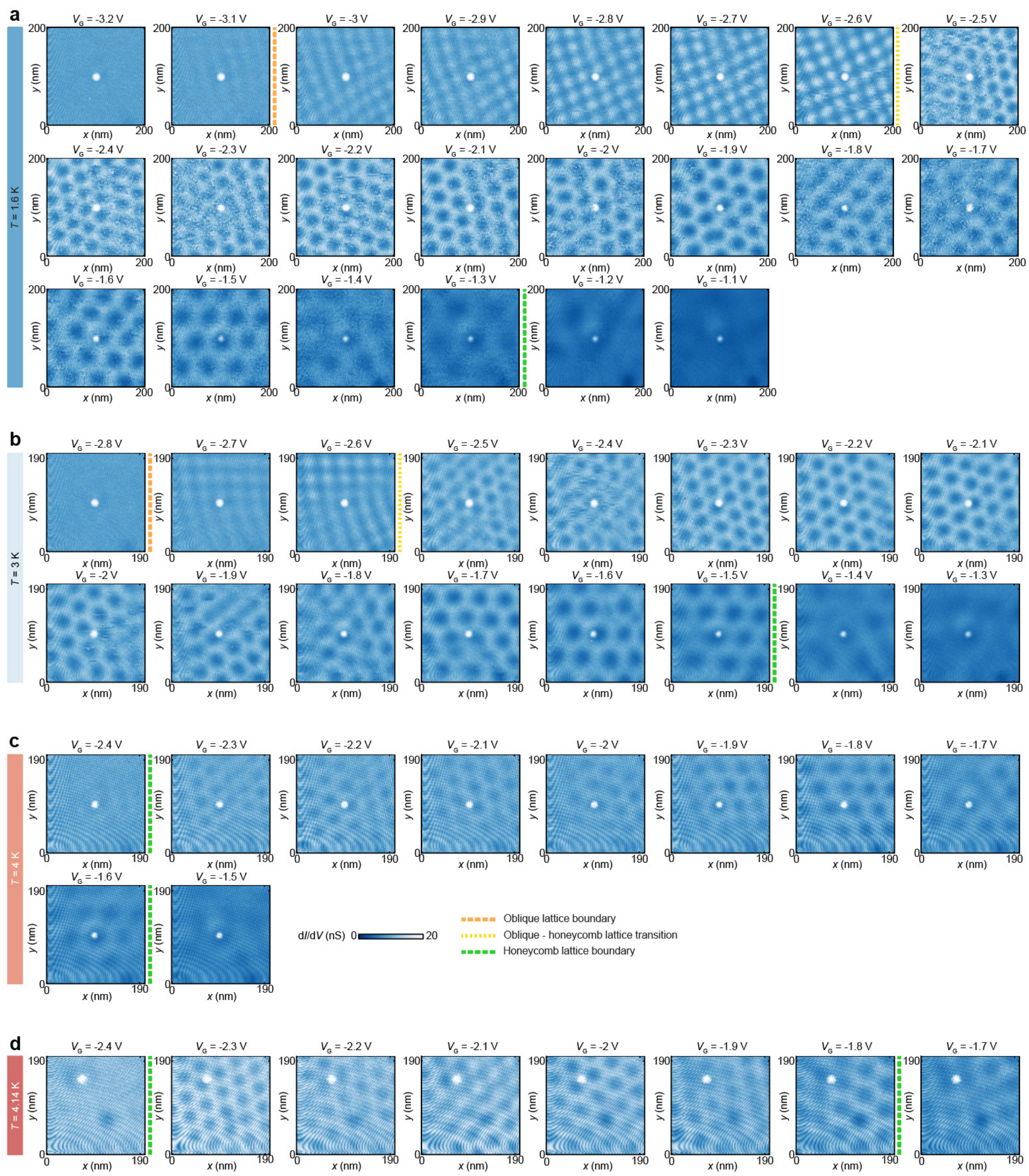

**Extended Data Fig. 8 | d*I*/d*V* maps as a function of gate voltages across the electron crystal regime at 1.6 K (a), 3 K (b), 4 K (c) and 4.14 K (d).** All images are acquired at sample bias $V_B$ = 2 mV. The tip heights are set by tunneling setpoints $V_B$ = 100 mV and $I_t$ = 5 nA, with lock-in modulation $V_{rms}$ = 0.4 mV. The 4 K data (**c**) are acquired with a tip-height offset $\Delta z$ = -20 pm. Phase boundaries are marked with colored dashed lines.

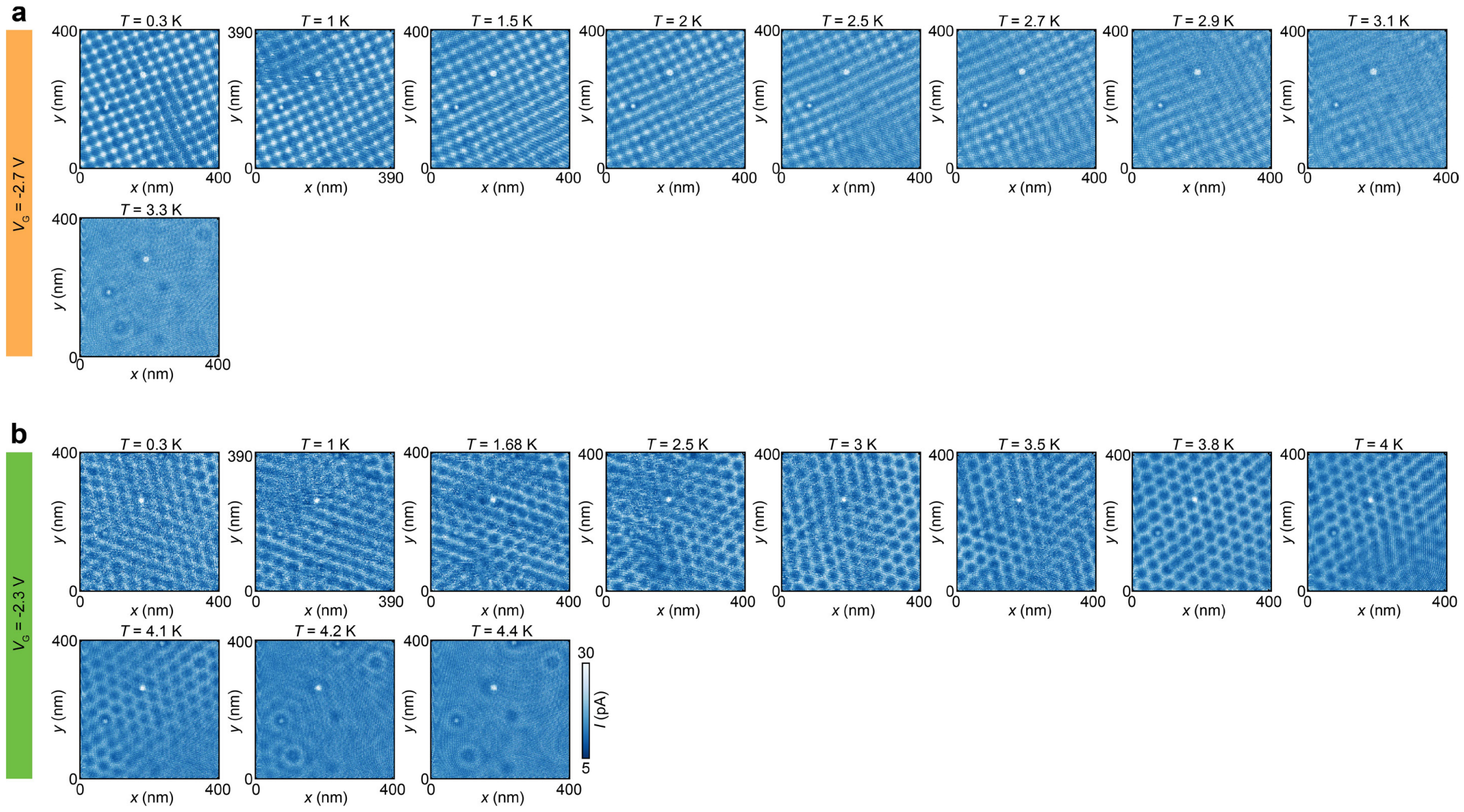

**Extended Data Fig. 9 | Tunneling current images showing the melting behavior of the oblique ($V_G$ = -2.7 V, a) and honeycomb ($V_G$ = -2.3 V, b) electron crystals.** The two crystal phases show distinct temperature evolutions. These maps are obtained at $V_B$ = 2 mV with tip height setpoint of $V_B$ = 100 mV and $I_t$ = 1 nA.

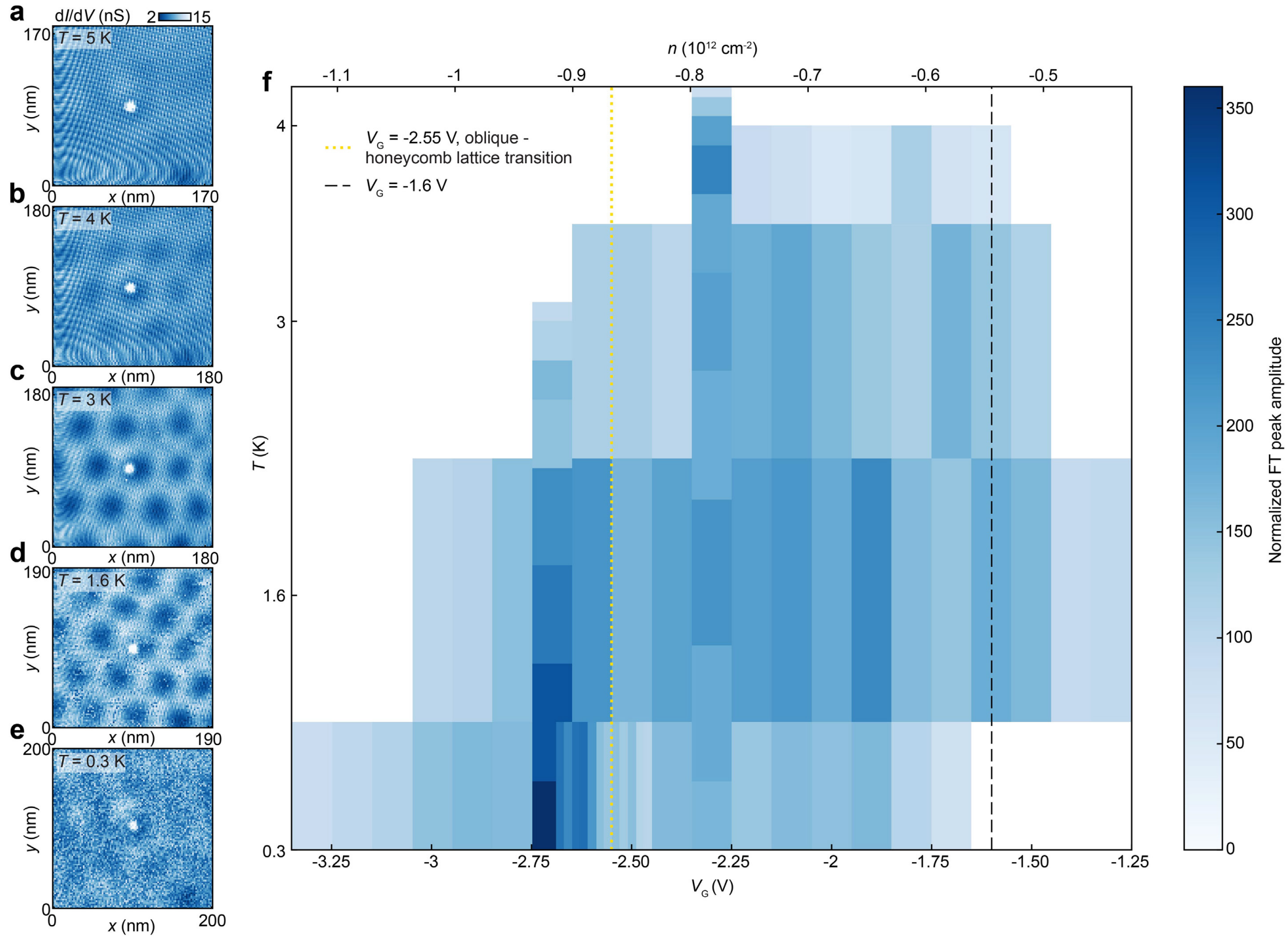

**Extended Data Fig. 10 | Further indication of Pomeranchuk-like behavior and temperature-dependent phase diagram quantified by the Fourier peak amplitude.** **a–e,** d$I$/d$V$ maps acquired at $V_G = -1.6$ V at temperatures of 5 K (**a**), 4 K (**b**), 3 K (**c**), 1.6 K (**d**) and 0.3 K (**e**). At $V_G$ = -1.6 V, although it lies within the inferred honeycomb crystal phase, the crystal pattern is visually nearly absent at the base temperature of 0.3 K (**e**). By contrast, upon increasing the temperature to 1.6 K and 3 K (**d, c**), pronounced crystal structures emerge before disappearing at 5 K (**a**). The crystal orientation also rotates slightly with heating, gradually evolving from approximately aligned with a rhombohedral graphene lattice direction to nearly parallel to the $x$ axis (also see Fig. 2k). This nonmonotonic temperature dependent evolution is not confined to $V_G$ = -1.6 V; it is widely seen in the honeycomb phase, in contrast to the monotonic temperature dependence of oblique lattice. **f**, Crystal Bragg peak amplitudes of the Fourier transformed tunneling current images at $V_B$ = 2 mV, normalized by the average tunneling current (see Methods for details), as a function of temperature and gate voltage. The oblique-to-honeycomb transition boundary and gate voltage corresponding to **a-e** ($V_G$ = -1.6 V) are indicated by the yellow and black dashed lines, respectively. This plot is obtained by analyzing data shown in Extended Data Fig. 5, 6, 8. Measurement conditions for **a–e**: imaging bias voltage, $V_B$ = 2 mV; setpoint conditions, $V_B$ = 100 mV and $I_t$ = 5 nA; lock-in modulation, $V_{rms}$ = 0.4 mV. The 4 K data (including **b** and corresponding raw data used in **f**) are acquired with a tip-height offset $\Delta z$ = -20 pm.

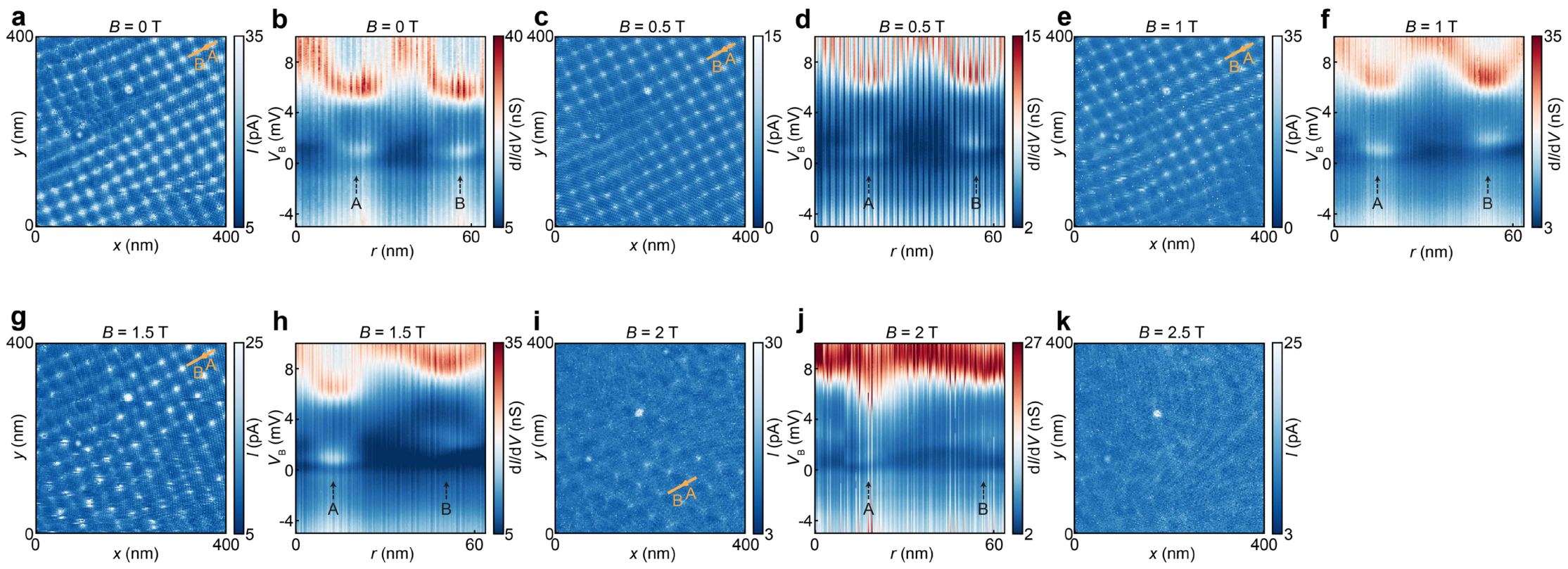

**Extended Data Fig. 11 | Magnetic field responses of the oblique crystal at $V_G$ = -2.7 V. a-k**, Real-space tunneling current maps acquired at $V_B$ = 2 mV, together with corresponding d$I$/d$V$ line spectra for magnetic fields ranging from $B$ = 0 to 2.5 T. The trajectories of the spectral linecuts are marked with orange arrows in the corresponding current maps. At zero field, the spectral peaks associated with the electron crystal near the Fermi level are nearly degenerate in energy at neighboring A and B lattice sites. With increasing magnetic field, orbital Zeeman coupling lifts the degeneracy, creating contrast between neighboring sites of the oblique lattice and resulting in a $\sqrt{2} \times \sqrt{2}$ superlattice reconstruction in real space (**c-i**). At a larger field of $B$ = 2.5 T (**k**), the electron crystal structure is destroyed and shows random fluctuations in real space. The d$I$/d$V$ linecut for **k** is therefore not included. Tip height setpoint: $V_B$ = 100 mV and $I_t$ = 1 nA.

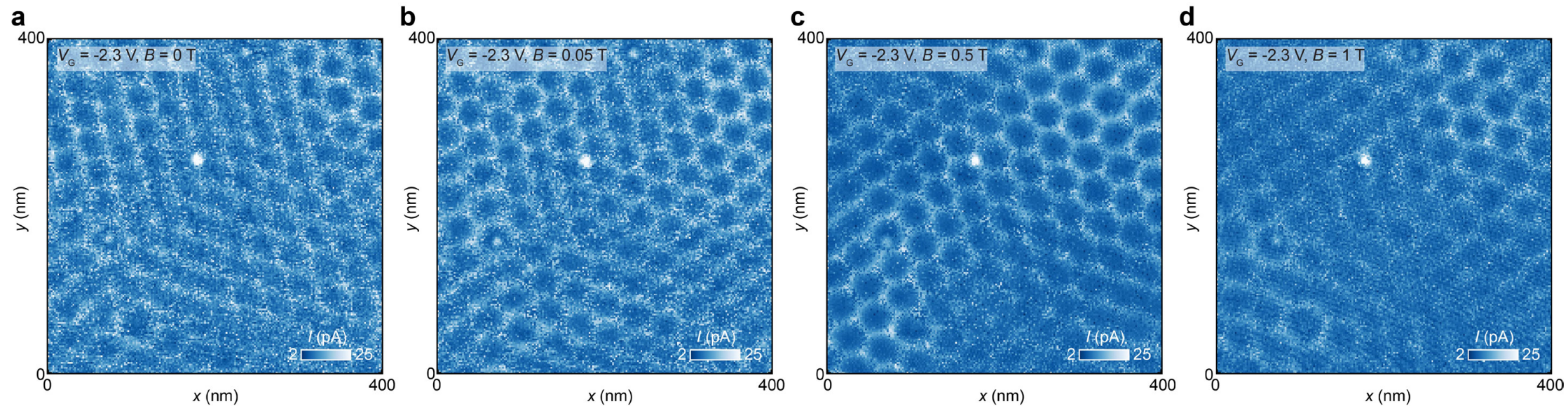

**Extended Data Fig. 12 | Magnetic responses of the honeycomb crystal at $V_G$ = -2.3 V.** Real-space tunneling current maps at $V_B$ = 2 mV under $B$ = 0 (**a**), 0.05 (**b**), 0.5 (**c**), and 1 T (**d**). To initialize the crystal state before acquiring **a**, we apply a cyclic gate-voltage sweep from $V_G$ = −2.3 V to 0 V and back to −2.3 V at a ramp rate of 2 V/s under zero magnetic field. The finite sweep rate kinetically hinders relaxation into a single ordered state, instead trapping the system in a metastable configuration. Consequently, the resulting crystal pattern exhibits blurred contrast with pronounced spatial inhomogeneity (**a**), indicative of a multidomain structure. Application of a small magnetic field ($B$ = 0.05 T, **b**) sharpens the crystal contrast and makes the pattern more uniform in space. Increasing the magnetic field to 0.5 T (**c**) preserves the overall symmetry of the honeycomb structure—unlike the oblique lattice (Fig. 4, Extended Data Fig. 11)—while enhancing crystal clarity and slightly increasing the lattice constant. At higher fields ($B$ = 1 T, **d**), crystalline order is locally suppressed, though the characteristic honeycomb lattice persists in regions where it remains resolved. Setpoint conditions: $V_B$ = 100 mV, $I_t$ = 1 nA.

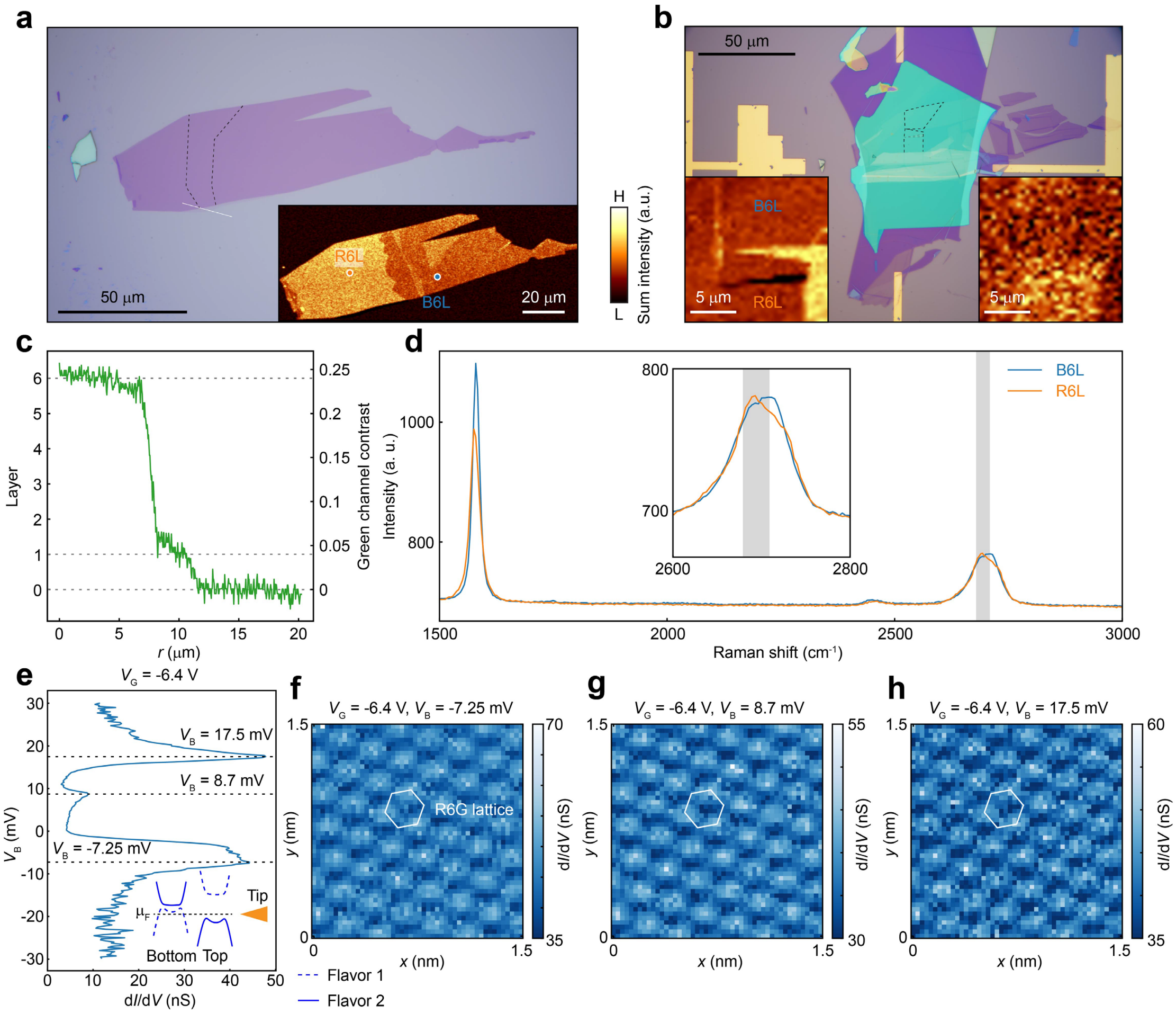

**Extended Data Fig. 13 | Device fabrication and characterization.** **a**, Optical microscope image of the hexalayer graphene flake used for device fabrication, containing domains of rhombohedral (R6L) and Bernal (B6L) stacking orders. The inset shows the stacking domains identified by Raman mapping of the 2D band, marked by orange (rhombohedral stacking) and blue (Bernal stacking) dots[54]. The R6L domain is cut and isolated along the black dashed line (Methods). H, high; L, low. **b**, Optical microscope image of finished R6L STM device. The outline of the isolated flake shown in **a** is indicated by the black dashed line. During stack fabrication, the upper region of the isolated R6G flake relaxed while the lower region remained in rhombohedral stacking, separated by cracks and wrinkles (grey dashed lines). The left inset shows a magnified Raman map of this region that clearly resolves the crack and wrinkle, which is obtained by integrating the G-band[54] intensity over a 30 cm$^{-1}$ window centered at 1588.7 cm$^{-1}$. The right inset shows the corresponding 2D-band intensity integrated over a 30 cm$^{-1}$ window centered at 2692.5 cm$^{-1}$, which distinguishes the two different stacking orders; the R6L region exhibits a higher intensity than the neighboring B6L regions. **c**, Linecut of the green-channel flat-field-corrected optical contrast along the white line in **a**. The optical contrast is approximately linear with graphene layer number under these imaging conditions and is calibrated here using our experimental parameters. **d**, Point Raman spectra acquired from the B6L and R6L regions (blue and orange points, respectively, in inset of **a**). The inset highlights a zoom-in view of the asymmetric 2D-band peak for R6L in contrast to B6L. The grey region corresponds to

the integration range of the inset in **a**. **e**–**h**, Point d$I$/d$V$ spectrum (**e**) and atomically resolved real-space d$I$/d$V$ maps acquired at the three spectral peaks (**f**–**h**) at $V_G$ = -6.4 V. Based on the band assignment discussed in Extended Data Fig. 3 and illustrated schematically in the inset of **e**, the peaks around $V_B$ = 8.7 mV and −7.25 mV correspond to the bottom- and top-layer valence band van Hove singularities, respectively. Generally, low energy states of rhombohedral graphene are expected to span active sublattices in opposite outmost layers (Fig. 1a). For rhombohedral hexalayer graphene specifically, however, the active sublattices of the top and bottom outer layers project onto the same $xy$ atomic positions, unlike in rhombohedral tetra- and penta-layer graphene (Fig. 1a). Consistent with this geometry, atomically resolved d$I$/d$V$ maps acquired at different bias voltages always display the same sublattice polarization (**f**–**h**). Similar behavior is observed over other gate voltages. Consequently, the layer polarization of these states and their possible interlayer correlation cannot be probed via atomic-scale imaging. This argument does not apply to rhombohedral trilayer graphene because its electronic wavefunctions extend much more to the middle layer[60,61]. Setpoint conditions: $V_B$ = 300 mV, $I_t$ = 100 pA, $\Delta z$ = -250 pm; $V_{rms}$ = 0.5 mV in **e**; $V_B$ = 100 mV, $I_t$ = 2 nA; $V_{rms}$ = 0.5 mV in **f**–**h**.